\documentclass[pdflatex,sn-mathphys-num]{sn-jnl}

\usepackage[british]{babel}
\usepackage{graphicx}
\usepackage{multirow}
\usepackage{amsmath,amssymb,amsfonts}
\usepackage{amsthm}
\usepackage{mathrsfs}
\usepackage[title]{appendix}
\usepackage{xcolor}
\usepackage{textcomp}
\usepackage{manyfoot}
\usepackage{booktabs}

\usepackage{xspace}
\usepackage{siunitx}
\usepackage{tikz}
\usepackage{pgf-pie}
\usepackage{colortbl}
\usepackage{subcaption}
\usepackage[section]{placeins}
\usetikzlibrary{arrows.meta, positioning, shapes.geometric, chains}

\newcommand{\tth}{$t\bar{t}H$\xspace}
\newcommand{\ttw}{$t\bar{t}W$\xspace}
\newcommand{\ttz}{$t\bar{t}Z$\xspace}
\newcommand{\MG}{\textsc{MadGraph5\_aMC@NLO}\xspace}
\newcommand{\Pythia}{\textsc{Pythia8}\xspace}
\newcommand{\Delphes}{\textsc{Delphes}\xspace}
\newcommand{\TREx}{\textsc{TRExFitter}\xspace}

\newcommand{\fs}[1]{\texttt{#1}}

\theoremstyle{thmstyleone}

\theoremstyle{thmstyletwo}

\theoremstyle{thmstylethree}

\usepackage[capitalize]{cleveref}

\begin{document}

\title[ML Methods for Event Selection]
{Benchmarking Machine Learning Architectures for ttH Multilepton Signal Sensitivity}

\author{\fnm{Luk\'{a}\v{s}} \sur{Vicen\'{i}k}, Andr\'{e} Sopczak, Oleksandr Shekhovtsov}
\email{lukas.vicenik@cern.ch}
\email{andre.sopczak@cern.ch}
\email{oleksandr.shekhovtsov@cvut.cz}

\affil{\orgdiv{\orgname{Czech Technical University in Prague}},
\orgaddress{\city{Prague}, \country{Czechia}}}

\abstract{
Statistical testing for signal discovery and signal-strength estimation in high-energy physics increasingly relies on machine-learning models trained on simulated data. We present a synthetic dataset for $t\bar t H$ multilepton
signal--background classification and perform a systematic evaluation
of machine-learning models ranging from the widely used XGBoost for
tabular data to LorentzNet, which processes events with built-in
Lorentz symmetry.
Existing studies often differ in feature definitions, training procedures, and evaluation metrics, making it difficult to isolate the impact of model architecture on performance. To address this, we apply standardized training and hyperparameter-optimization procedures, construct a controlled hierarchy of feature sets, and perform a comprehensive comparison of the models.
In addition to the analysis-driven metric of signal-strength
uncertainty, we report the commonly used ROC AUC metric and show that under realistically weighted training, it correlates well with the
uncertainty-based ranking of models.
We further investigate model performance as a function of input feature set, training-set size, and the choice between channel-specific and unified multi-channel training.
Our results establish the superiority of Particle Transformer and LorentzNet within the considered setup. We also identify potential avenues for further improvement, including more expressive architectures, joint training across additional analysis channels, and larger simulated datasets. Although this study focuses on a specific Higgs-boson analysis, we expect the main conclusions to generalize to a broader class of searches and measurements at the LHC and HL-LHC.

}


\maketitle

\section{Introduction}\label{sec:intro}

Machine learning has become central to signal-background separation
in high-energy physics (HEP), yet event-level classification in phenomenological studies and many experimental analyses has been predominantly depending on a set of basic machine learning approaches: engineer
high-level kinematic variables and train a Gradient Boosted Decision Tree (GBDT) on the resulting tabular representation
\cite{Mondal:2024nsa, Liu:2026spin2, Benbrik:2026hvq, ATLAS:2025wwy}.
This workflow involves two rarely questioned design choices: manual
feature engineering, which constructs physics-aware discriminative
features but discards low-level information, and a fixed-slot
representation introduced by the column ordering of the input dataset. Both are architectural
conveniences, not physical requirements. This data representation approach is known to work particularly well with GBDTs.

The natural alternative to the described tabular approach is to represent the proton-proton (pp) collision event constituents and process their elementary kinematics properties with transformations that are designed to be permutationally invariant (e.g. message passing graph neural networks or
attention models~\cite{Bronstein:2021gdl}). Such an approach allows viewing the constituents and their
feature vectors individually, a design possibility unavailable to
GBDTs and multilayer perceptron (MLP), and enables geometric properties
such as permutation, rotation, or Lorentz invariance to be natively
enforced by construction, rather than approximated or ignored as in
fixed-slot tabular representations. Such geometric-based architectures have demonstrated substantial gains
in jet tagging~\cite{Qu:2019gqs,Qu:2022mxj,Gong:2022ghc}, but
whether these gains transfer to event-level classification remains an
open question.

Beyond architectural comparisons, the choice of evaluation metric
matters. ROC AUC measures global ranking quality independently of
any statistical model or luminosity assumption, while profile
likelihood sensitivity $\varrho_\mu$ folds in the full statistical
analysis (luminosity, bin yields, and MC statistical uncertainties) and is therefore more directly connected to analysis impact. The
two metrics are not guaranteed to agree in general, so reporting both
is necessary for a complete picture of classifier performance.

We address the gaps mentioned above with a controlled benchmark framework. Six classifier architectures are evaluated on \tth multilepton signal--background separation: XGBoost~\cite{Chen:2016xgboost}, MLP, Message Passing Neural Network (MPNN)~\cite{Gilmer:2017mpnn}, MPNN with Rotary Position Encoding (MPNN+RoPE), Particle Transformer (ParT)~\cite{Qu:2022mxj}, and LorentzNet~\cite{Gong:2022ghc}. These span flat tabular models, generic graph networks, and architectures with explicit physical symmetry constraints, and are evaluated at $\sqrt{s}=14$~TeV, corresponding to the High Luminosity LHC (HL-LHC), the upgraded operating phase expected to deliver an integrated luminosity of up to 3000~fb$^{-1}$~\cite{BejarAlonso:2020hlhlhc}, under a four-level feature hierarchy, using relative profile likelihood uncertainty $\varrho_\mu$ as the key performance indicator. No public benchmark dataset exists for event-level \tth\ multilepton classification; we therefore generate a dedicated simulation sample designed to reflect HL-LHC conditions.

We show that especially graph-based architectures with incorporated symmetry constraints outperform flat tabular models substantially. ParT and LorentzNet achieve the highest ROC AUC among all six architectures, with the two nearly tied: ParT at the \fs{full} feature level, LorentzNet at \fs{object}, the highest level at which its Lorentz-equivariant design remains applicable. LorentzNet additionally achieves the best $\varrho_\mu$ of all six architectures.
Encoding the detector's azimuthal symmetry alone (via single-frequency rotary position encoding in an otherwise standard MPNN) yields a consistent gain over unconstrained baselines across all feature levels and data fractions, narrowing the gap to architectures with explicit physics-motivated inductive biases (ParT, LorentzNet) noticeably. This suggests that symmetry constraints are, alongside architectural sophistication, a primary driver of improvement over unconstrained models, though a meaningful performance gap to the more complex architectures remains. We further show that ROC AUC and $\varrho_\mu$ correlate well and preserve the same architecture ranking, indicating that improvements in classifier score quality translate consistently into improvements in profile likelihood sensitivity for this benchmark.
Finally, joint multi-channel
training improves per-channel performance, with the
largest gains for graph-based architectures in data-scarce
channels, supporting a foundation model approach to
signal--background separation. In addition to the benchmark results, we
release a dedicated \tth\ multilepton simulation sample of 593{,}390 pre-selected events across six analysis channels at $\sqrt{s}=14$~TeV,
the first public dataset of this kind targeting HL-LHC conditions.

\section{Related work}\label{sec:related}
Graph-based architectures have established strong benchmarks in jet
tagging, with ParticleNet~\cite{Qu:2019gqs}, ParT~\cite{Qu:2022mxj},
and LorentzNet~\cite{Gong:2022ghc} evaluated on standardised
top-tagging and quark-gluon datasets. Whether the same architectural
advantages hold at the event level is less well established: the
input is a heterogeneous mixture of object types with variable
multiplicity, and discriminating correlations span the full event
topology rather than a single jet cone.

Nevertheless, graph-based models have begun appearing in real
experimental analyses at the event level. The ATLAS \tth\ multilepton
measurement~\cite{ATLAS:2025eua} employs a GNN for
signal--background separation across the same six channels studied
here, using particle four-momenta as node features and angular
separations as edge features. The ATLAS four-top-quark
observation~\cite{ATLAS:2023ajo} similarly uses a GNN in a
multilepton final state, reporting a 10\% improvement in expected
significance over a GBDT baseline. These results confirm that
event-level graph models are sufficiently mature for deployment in
LHC measurements, but provide no systematic comparison of
architectures or evaluation of physical symmetry properties.

On the benchmarking side, Pfeffer et al.~\cite{Pfeffer:2024csbs}
perform a careful controlled comparison of GraphConv-based GNNs
against fully connected deep neural networks (DNNs) for event-level $t\bar{t}$+X
classification, explicitly matching architectures in terms of
trainable parameters and input information. They find that GNN
advantages can be traced to relational information between objects
rather than to the graph structure itself, and are most pronounced
for hierarchically structured data. Onyisi et
al.~\cite{Onyisi:2022vyu} compare point cloud strategies for
collider event classification across several graph-based
architectures, and Builtjes et al.~\cite{Builtjes:2025spi} compare
attention and graph-based architectures with a focus on transformer
variants. None of these studies investigates the impact of physical symmetry constraints
on classifier performance, uses profile likelihood sensitivity as an evaluation metric, or 
performs a controlled feature hierarchy ablation.

Both Pfeffer et al.\ and Onyisi et al.\ generate private simulation
samples with no public release; the only public event-level analogue
is the DarkMachines four-tops dataset~\cite{Aarrestad:2021oeb} used
by Builtjes et al., which covers a single signal region at
leading order (LO) in QCD at $\sqrt{s}=13$~TeV and does not support
multi-channel or joint-versus-dedicated training comparisons. No
public dataset exists for event-level \tth multilepton
classification at HL-LHC conditions; we therefore generate a
dedicated NLO sample, described in \cref{sec:dataset}.

\section{Multichannel Simulated Dataset}\label{sec:dataset}

We generated a new public dataset for event-level \tth multilepton classification at HL-LHC conditions, with the signal and background processes, channel definitions, and object selection designed to closely follow the ATLAS analysis~\cite{ATLAS:2025eua}. Several deliberate simplifications are made to isolate architectural differences: detector response is modelled with fast simulation rather than full \textsc{Geant4}.

The dataset, containing selected events across six mutually exclusive analysis channels, with per-event four-momenta, object types, charges, $b$-tag scores, and process labels, will be released alongside this paper under the CC-BY-4.0 licence, together with the full simulation configuration files and selection code, to support
reproducibility and future benchmarking studies.

\subsection{Simulation chain}
Events are generated at $\sqrt{s}=14$~TeV at next-to-leading order (NLO) in quantum chromodynamics (QCD) with \MG~\cite{Alwall:2014hca} using
the NNPDF~3.1 NNLO PDF set with $\alpha_s(M_Z)=0.118$ and four active
flavours (\texttt{NNPDF31\_nnlo\_as\_0118\_nf\_4}~\cite{NNPDF:2017mvq}),
interfaced with \Pythia~\cite{Sjostrand:2014zea} for parton showering
and hadronisation. Detector response is modelled with
\Delphes~3.5 using the standard ATLAS detector
card~\cite{deFavereau:2013fsa}. 

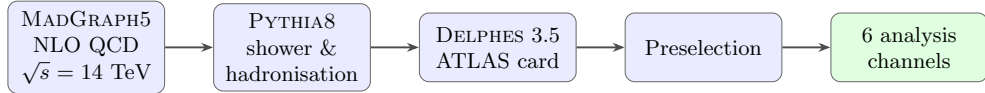
\begin{figure}[!ht]
\centering
\resizebox{\textwidth}{!}{
\begin{tikzpicture}[
  node distance   = 0.5cm and 0.7cm,
  box/.style      = {rectangle, rounded corners=4pt, draw=black!60,
                     fill=blue!8, text width=2.0cm, align=center,
                     font=\small, minimum height=1.0cm, inner sep=4pt},
  finalbox/.style = {rectangle, rounded corners=4pt, draw=black!60,
                     fill=green!12, text width=2.0cm, align=center,
                     font=\small, minimum height=1.0cm, inner sep=4pt},
  arr/.style      = {-{Stealth[length=5pt]}, thick, black!70}
]
  \node[box]      (mg) {\small \textsc{MadGraph5}\\\small NLO QCD\\$\sqrt{s}=14$~TeV};
  \node[box,      right=of mg] (py) {\textsc{Pythia8}\\\small shower \&\\hadronisation};
  \node[box,      right=of py] (dl) {\textsc{Delphes}~3.5\\\small ATLAS card};
  \node[box,      right=of dl] (ps) {Preselection};
  \node[finalbox, right=of ps] (ch) {6 analysis\\channels};
  \draw[arr] (mg) -- (py);
  \draw[arr] (py) -- (dl);
  \draw[arr] (dl) -- (ps);
  \draw[arr] (ps) -- (ch);
\end{tikzpicture}}\\[5pt]
\caption{Simulation and reconstruction chain. Events are generated with
\MG, showered and hadronised with \Pythia, passed through the
\Delphes\ fast detector simulation with the ATLAS card, and
categorised into six mutually exclusive analysis channels after preselection.}
\label{fig:sim_chain}
\end{figure}

\subsection{Simulated processes and dataset composition}
Three processes are simulated: signal \tth (inclusive Higgs decay) and
prompt backgrounds \ttw and \ttz. These represent the dominant sources
of prompt multileptons in the relevant phase space and are well-modelled
by simulation, making them suitable for an ML benchmark.
Non-prompt lepton contributions are not specified separately because they are included in the simulation and not determined by a data-driven method. Since all architectures are
evaluated on the same simulated dataset, this does not affect the
relative ranking of classifiers. Cross-sections $\sigma$, generated event counts $N_{\rm gen}$,
preselection yields, and efficiencies are given in \cref{tab:dataset}. The total dataset comprises 593{,}390 events.

\begin{table}[t]
\caption{Cross-sections at $\sqrt{s}=14$~TeV~\cite{LHCHiggsCrossSectionWorkingGroup:2016ypw}, generated event counts,
events surviving preselection, and preselection efficiencies. All three
processes enter the profile likelihood fit.}
\label{tab:dataset}
\begin{tabular}{@{}lrrrr@{}}
\toprule
Process & $\sigma$ [pb] & $N_\mathrm{gen}$ & $N_\mathrm{sel}$ & 
$\varepsilon$ [\%] \\
\midrule
\tth\ (signal) & $0.599$ & 32{,}900{,}000 & 207{,}051 & 0.63 \\
\ttw           & $0.824$ & 28{,}400{,}000 & 326{,}413 & 1.15 \\
\ttz           & $0.998$ & 13{,}000{,}000 &  59{,}926 & 0.46 \\
\midrule
Total          &         &                & 593{,}390 &      \\
\botrule
\end{tabular}
\end{table}

\subsection{Object reconstruction and preselection}\label{subsec:presel}
Object reconstruction, the process of building physics objects such as jets, $b$-tagged jets, leptons, and tau candidates from the raw detector response, is performed within the Delphes fast simulation. Jets are clustered using the anti-$k_T$ algorithm~\cite{Cacciari:2008gp} with $R=0.6$ and $p_T>20$~GeV;
$b$-jets are identified at the 70\% efficiency working
point~\cite{ATL-PHYS-PUB-2015-022}. Tau candidates are reconstructed
via track counting~\cite{ATL-PHYS-PUB-2015-045}.
Up to eight jets, four leptons, and two tau candidates are retained per
event. Events are categorised into six mutually
exclusive analysis channels following the ATLAS
definitions~\cite{ATLAS:2025eua}: $2\ell$SS$0\tau$, $3\ell0\tau$,
$4\ell$, $2\ell$SS$1\tau$, $1\ell2\tau$, and $2\ell$$2\tau$. Full
selection specifications are given in \cref{app:selection}.

Per-channel event counts and their fractional composition are shown in \cref{fig:channel_counts}. The sample is strongly dominated by the $2\ell$SS$0\tau$ channel 
(69.7\% of all selected events), reflecting the higher branching 
fraction and looser selection requirements of same-sign dilepton 
final states; the $4\ell$ and $2\ell$$2\tau$ channels
together contribute only 2.2\%, a class imbalance that directly
motivates the dedicated versus multi-channel training comparison in \cref{sec:channels}.

\begin{figure}[t]
\centering
\captionof{table}{Per-channel event counts after preselection, broken down by process.}
\begin{tabular}{@{}lrrrrrrr@{}}
\toprule
Channel & $N_{t\bar{t}H}$ & $N_{t\bar{t}W}$ & $N_{t\bar{t}Z}$ & $N_\mathrm{sel}$ & Fraction [\%] & $f_\mathrm{sig}$ [\%] \\
\midrule
$2\ell$SS$0\tau$ & 113{,}113 & 264{,}487 & 36{,}140 & 413{,}740 & 69.7 & 27.3 \\
$3\ell0\tau$     &  42{,}062 &  45{,}635 &  7{,}532 &  95{,}229 & 16.0 & 44.2 \\
$1\ell2\tau$     &  30{,}633 &   8{,}220 &  6{,}135 &  44{,}988 &  7.6 & 68.1 \\
$2\ell$SS$1\tau$ &  14{,}993 &   7{,}728 &  3{,}226 &  25{,}947 &  4.4 & 57.8 \\
$4\ell$          &   3{,}271 &      29   &  6{,}439 &   9{,}739 &  1.6 & 33.6 \\
$2\ell$$2\tau$ &   2{,}979 &     314   &    454   &   3{,}747 &  0.6 & 79.5 \\
\midrule
Total            & 207{,}051 & 326{,}413 & 59{,}926 & 593{,}390 & 100  & 34.9 \\
\botrule
\end{tabular}

\vspace{0.5cm}

\begin{minipage}{0.48\textwidth}
\centering
\begin{tikzpicture}
\pie[radius=2.0, text=legend, before number=, after number=\%,
    color={blue!55,red!50,green!50,orange!55,purple!45,cyan!50}]{
    69.7/$2\ell$SS$0\tau$,
    16.0/$3\ell0\tau$,
     7.6/$1\ell2\tau$,
     4.4/$2\ell$SS$1\tau$,
     1.6/$4\ell$,
     0.6/$2\ell$$2\tau$
}
\end{tikzpicture}
\end{minipage}%
\hfill
\begin{minipage}{0.48\textwidth}
\centering
\begin{tikzpicture}[x=0.4mm, y=0.55cm]
  \definecolor{colH}{RGB}{70,130,180}
  \definecolor{colW}{RGB}{205,92,92}
  \definecolor{colZ}{RGB}{60,160,60}
  \def\bh{0.32}
  \def\W{100}

  \fill[colH](0,0-\bh)rectangle(79.5,0+\bh);
  \fill[colW](79.5,0-\bh)rectangle(87.9,0+\bh);
  \fill[colZ](87.9,0-\bh)rectangle(100,0+\bh);
  \node[anchor=east,font=\small]at(-1,0){$2\ell$$2\tau$};

  \fill[colH](0,1-\bh)rectangle(33.6,1+\bh);
  \fill[colW](33.6,1-\bh)rectangle(33.9,1+\bh);
  \fill[colZ](33.9,1-\bh)rectangle(100,1+\bh);
  \node[anchor=east,font=\small]at(-1,1){$4\ell$};

  \fill[colH](0,2-\bh)rectangle(57.8,2+\bh);
  \fill[colW](57.8,2-\bh)rectangle(87.6,2+\bh);
  \fill[colZ](87.6,2-\bh)rectangle(100,2+\bh);
  \node[anchor=east,font=\small]at(-1,2){$2\ell$SS$1\tau$};

  \fill[colH](0,3-\bh)rectangle(68.1,3+\bh);
  \fill[colW](68.1,3-\bh)rectangle(86.4,3+\bh);
  \fill[colZ](86.4,3-\bh)rectangle(100,3+\bh);
  \node[anchor=east,font=\small]at(-1,3){$1\ell2\tau$};

  \fill[colH](0,4-\bh)rectangle(44.2,4+\bh);
  \fill[colW](44.2,4-\bh)rectangle(92.1,4+\bh);
  \fill[colZ](92.1,4-\bh)rectangle(100,4+\bh);
  \node[anchor=east,font=\small]at(-1,4){$3\ell0\tau$};

  \fill[colH](0,5-\bh)rectangle(27.3,5+\bh);
  \fill[colW](27.3,5-\bh)rectangle(91.3,5+\bh);
  \fill[colZ](91.3,5-\bh)rectangle(100,5+\bh);
  \node[anchor=east,font=\small]at(-1,5){$2\ell$SS$0\tau$};

  \draw[->](0,-0.55)--(105,-0.55);
  \foreach \x in {0,25,50,75,100}{
    \draw(\x,-0.55)--(\x,-0.7);
    \node[below,font=\small]at(\x,-0.7){\x\%};
  }

  \fill[colH](20,5.6)rectangle(30,5.9);
  \node[anchor=west,font=\small]at(31,5.75){$t\bar{t}H$};
  \fill[colW](50,5.6)rectangle(60,5.9);
  \node[anchor=west,font=\small]at(61,5.75){$t\bar{t}W$};
  \fill[colZ](80,5.6)rectangle(90,5.9);
  \node[anchor=west,font=\small]at(91,5.75){$t\bar{t}Z$};
\end{tikzpicture}
\end{minipage}

\captionof{figure}{The pie chart shows the overall channel composition. The stacked bar chart shows the per-channel class composition: $2\ell$$2\tau$ and
$1\ell2\tau$ are signal-rich while $2\ell$SS$0\tau$ and $3\ell0\tau$ are
background-dominated, driven by large $t\bar{t}W$ contributions. The negligible $t\bar{t}W$ yield in the $4\ell$ channel (29 events) is
physically expected, as $t\bar{t}W$ can produce at most three prompt leptons at tree level.}
\label{fig:channel_counts}
\end{figure}

\subsection{Event weights}\label{sec:weights}

The generated dataset does not automatically reflect the naturally
expected distribution of collision events: the number of events
simulated per process $N_{\mathrm{gen},p}$ is an arbitrary generation
choice, not proportional to the physical process rate. At a given
integrated luminosity $\mathscr{L}$, the expected number of events
for process $p$ surviving preselection is
\begin{equation}
N_{\mathrm{exp},p} = \mathscr{L} \cdot \sigma_p \cdot \varepsilon_p,
\label{eq:expected}
\end{equation}
where $\sigma_p$ is the process cross-section.
To recover this physical mixture independently of $\mathscr{L}$, each
selected event from process $p$ is assigned a weight
\begin{equation}
w_e = \frac{\sigma_{\mathrm{proc}(e)}}{N_{\mathrm{gen},\mathrm{proc}(e)}},
\label{eq:weights}
\end{equation}
such that the weighted sample composition matches the naturally expected
process rates. These weights are per-process constants: all selected
events from the same process receive the same weight, since no
per-event kinematic reweighting is applied. No additional reweighting
is applied to balance signal and background; the sample retains the
natural signal-to-background ratio. The use of these weights in the analysis pipeline is described in \cref{sec:hpo}.

\subsection{Event content}\label{sec:event_content}

Each simulated event, after preselection and channel categorisation
(\cref{subsec:presel}), is stored as a variable-length
collection of reconstructed physics objects together with event-level
metadata, independently of how that content is later encoded for a
given architecture or organised into the feature hierarchy
(\cref{sec:representation,sec:feature_hierarchy}).

An event contains up to four leptons (electrons or muons), two tau
candidates, and eight jets, for a maximum of $N\leq14$ objects; the
actual number present varies event-by-event according to the physics
process and analysis channel. Each object carries an object-type
label (\texttt{electron}, \texttt{muon}, \texttt{tau}, \texttt{jet})
and a four-momentum. Leptons and $\tau_\mathrm{had}$ candidates
additionally carry an electric charge $q\in\{-1,+1\}$; jets carry a
$b$-tag score, a continuous discriminant of the likelihood that the
jet originated from a $b$-quark. Slots not filled by a reconstructed
object in a given event are recorded as absent.

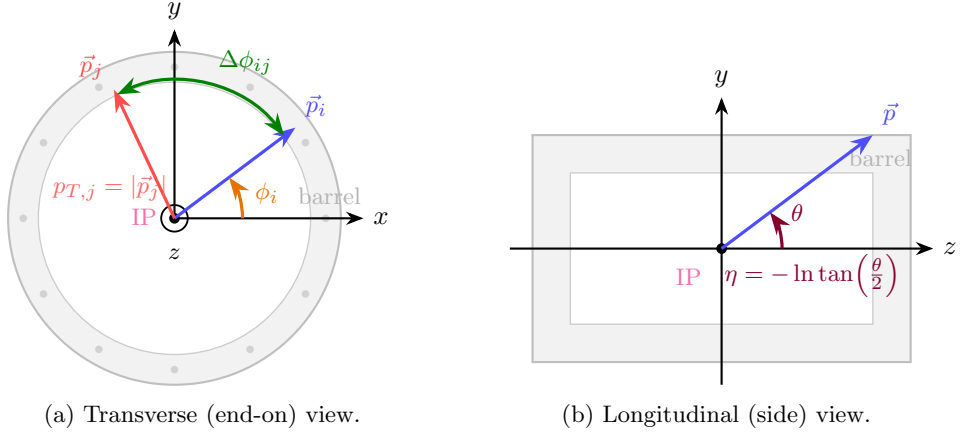
\begin{figure}[t]
\centering
\resizebox{\textwidth}{!}{%
\begin{minipage}{\textwidth}
\centering
\begin{subfigure}{0.54\textwidth}
\centering
\begin{tikzpicture}[scale=1.0]
  \draw[fill=gray!10, draw=gray!50, line width=0.8pt]
    (0,0) circle (2.2cm);
  \draw[fill=white, draw=gray!40, line width=0.5pt]
    (0,0) circle (1.8cm);
  \node[gray!60, font=\small] at (2.05, 0.3) {barrel};
  \foreach \a in {0,30,...,330}
    \fill[gray!35] ({2.0*cos(\a)},{2.0*sin(\a)}) circle (0.05cm);
  \fill[black] (0,0) circle (2pt);
  \draw[black, line width=0.7pt] (0,0) circle (5pt);
  \node[font=\small, anchor=north, yshift=-8pt] at (0,0) {$z$};
  \node[font=\small, magenta!70, anchor=south, yshift=8pt] at (-0.4,-0.5) {\small IP};
  \draw[-{Stealth}, thick] (0,0) -- (2.5,0) node[right] {$x$};
  \draw[-{Stealth}, thick] (0,0) -- (0,2.5) node[above] {$y$};
  \draw[-{Stealth}, very thick, blue!70] (0,0) -- (1.6, 1.2)
    node[above right, font=\small] {$\vec{p}_i$};
  \draw[-{Stealth}, orange!90!black, very thick]
    (0.9,0) arc(0:36.87:0.9cm)
    node[right, xshift=6pt, yshift=-6pt, font=\small] {$\phi_i$};
  \draw[-{Stealth}, very thick, red!70] (0,0) -- (-0.8, 1.7)
    node[above left, font=\small] {$\vec{p}_j$};
  \node[red!60, font=\small] at (-0.85, 0.40) {$p_{T,j}=|\vec{p}_j|$};
  \draw[{Stealth}-{Stealth}, green!50!black, very thick]
    (1.45, 1.09) arc(36.87:115.3:1.8cm)
    node[midway, above right, font=\small] {$\Delta\phi_{ij}$};
\end{tikzpicture}
\caption{Transverse (end-on) view.}
\end{subfigure}
\hfill
\begin{subfigure}{0.42\textwidth}
\centering
\begin{tikzpicture}[scale=1.0]
  \draw[fill=gray!10, draw=gray!50, line width=0.8pt]
    (-2.5,-1.5) rectangle (2.5,1.5);
  \draw[fill=white, draw=gray!40, line width=0.5pt]
    (-2.0,-1.0) rectangle (2.0,1.0);
  \node[gray!60, font=\small] at (2.1, 1.2) {barrel};
  \draw[-{Stealth}, thick] (-2.8,0) -- (2.8,0)
    node[right] {$z$};
  \draw[-{Stealth}, thick] (0,-1.8) -- (0,2.0)
    node[above] {$y$};
  \fill[black] (0,0) circle (2pt);
  \node[font=\small, magenta!70, anchor=north east, xshift=-4pt, yshift=-4pt]
    at (0,0) {\small IP};
  \draw[-{Stealth}, very thick, blue!70] (0,0) -- (2.0, 1.5)
    node[above right, font=\small] {$\vec{p}$};
  \draw[-{Stealth}, purple!70!black, very thick]
    (0.8,0) arc(0:36.87:0.8cm)
    node[right, xshift=4pt, font=\small] {$\theta$};
  \node[font=\small, purple!70!black] at (1.2, -0.35)
    {$\eta = -\ln\tan\!\left(\tfrac{\theta}{2}\right)$};
\end{tikzpicture}
\caption{Longitudinal (side) view.}
\end{subfigure}
\end{minipage}%
}
\caption{ATLAS detector coordinate system. \emph{Left}: end-on
($x$-$y$) view with the beam axis $z$ pointing toward the reader
($\odot$). Each particle is characterised by its transverse momentum
$p_T = \sqrt{p_x^2 + p_y^2}$, the momentum component in the
$x$-$y$ plane, and azimuthal angle $\phi$, measured from the
$x$-axis; only the difference $\Delta\phi_{ij}$ between particles
$i$ and $j$ is physically meaningful. \emph{Right}: longitudinal
($y$-$z$) view showing the polar angle $\theta$ between the particle
trajectory and the beam axis $z$, and the pseudorapidity $\eta =
-\ln\tan(\theta/2)$, which is the standard longitudinal coordinate
used in place of $\theta$ at hadron colliders.}
\label{fig:coordinates}
\end{figure}

Particle kinematics are expressed in two equivalent coordinate
systems, illustrated in \cref{fig:coordinates}. The Cartesian
four-momentum $(E, p_x, p_y, p_z)$ is Lorentz-covariant and natural
for computing invariant masses and Minkowski inner products, and is
stored as the primitive representation from which the detector
coordinates are derived. The detector coordinates $(p_T, \eta, \phi,
m)$ are adapted to the cylindrical geometry of the ATLAS detector: as
shown in \cref{fig:coordinates} (left), $p_T = \sqrt{p_x^2 +
p_y^2}$ is the transverse momentum magnitude in the $x$-$y$ plane and
$\phi = \arctan(p_y/p_x)$ is the azimuthal angle measured from the
$x$-axis; as shown in \cref{fig:coordinates} (right), $\theta$
is the polar angle between the particle trajectory and the beam axis
$z$, and $\eta = -\ln\tan(\theta/2)$ is the pseudorapidity, the
standard longitudinal coordinate at hadron colliders. The fourth
detector coordinate, $m = \sqrt{E^2 - p_x^2 - p_y^2 - p_z^2}$, is the
invariant mass of the particle, computed directly from the Cartesian
four-momentum. A simultaneous rotation $\phi_i \to \phi_i +
\Delta\phi_0$ of all particles leaves the physics invariant.

Beyond per-object content, each event carries six derived event-level
quantities: missing transverse energy $E_{\mathrm{T}}^{\mathrm{miss}}$
(MET), the scalar transverse energy sum $\rm HT$, and the object
multiplicities $n_\mathrm{jets}$, $n_\mathrm{bjets}$,
$n_\mathrm{leptons}$, $n_\mathrm{taus}$. $E_{\mathrm{T}}^{\mathrm{miss}}$
and $\rm HT$ are recomputed from the selected objects rather than
taken directly from the Delphes output, ensuring consistency with the
particle-level feature set. Each event additionally carries the
parent process ($t\bar tH$, $t\bar tW$, or $t\bar tZ$), the analysis
channel, and the event weight~\eqref{eq:weights}, used respectively
for labelling, categorisation, and training/evaluation weighting
rather than as classifier input.

\section{Architectures and their Symmetries}\label{sec:arch_sym}
This section introduces the six benchmarked architectures: flat models (MLP, XGBoost) and graph models (MPNN, MPNN+RoPE, ParT, LorentzNet), spanning the spectrum of physical symmetry constraints that reflects the geometry of the ATLAS detector. How each group encodes the event content of \cref{sec:event_content} as a model input is described in \cref{sec:representation}.
The LHC beam axis breaks full Lorentz invariance in two ways: finite
detector acceptance in pseudorapidity $\eta$ makes only rotations
about the beam axis leave the detected particles unchanged, and the
event-varying parton momentum fractions $x_1, x_2$ of the colliding
protons leave the boost to the partonic
centre-of-mass frame undetermined and therefore unusable as a
symmetry. Together these reduce the exact symmetry group
$\mathrm{SO}(3,1)$ to its cylindrical subgroup: azimuthal rotation
about the beam axis, $\mathrm{SO}(2)$, simultaneous rotation of all
objects by a common angle $\Delta\phi_0$, remains an exact symmetry of
both the physics process and the ideal detector, since neither the
hard scattering nor the cylindrical detector geometry defines a
preferred azimuthal direction. Any dependence of the classifier on the
absolute event orientation $\phi$ therefore reflects simulation
structure rather than genuine discriminating physics, and will not
generalise from simulation to data.

\paragraph{XGBoost}
XGBoost learns an additive ensemble of
decision trees: at each step, select one feature and branch on its value compared to a learned threshold, until a leaf node is reached that stores a score. XGBoost captures high-order feature
interactions through split combinations but cannot learn
permutation-invariant representations from the fixed-position flat
feature layout. As the standard approach in current experimental and
phenomenological analyses, XGBoost serves as the de facto
analysis baseline against which the remaining five architectures are
compared.

\paragraph{MLP}
The multilayer perceptron is the simplest of the six architectures: a
sequence of fully-connected layers applied to a fixed-length input
vector. As a flat model, event objects appear as fixed-position columns
in a $p_T$-sorted feature vector, encoding no symmetry constraints.
The MLP is therefore sensitive to both object permutation and absolute
azimuthal orientation $\phi$, and serves as the flat-architecture
reference point against which graph-based models are compared.

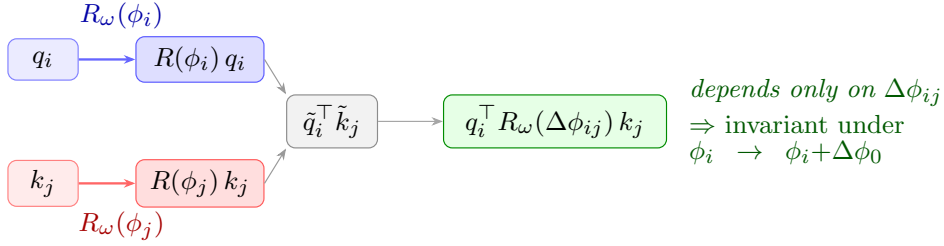
\begin{figure}[!ht]
\centering
\resizebox{\textwidth}{!}{%
\begin{tikzpicture}[font=\small]
\node[draw=blue!50, fill=blue!8, rounded corners=3pt,
      minimum width=0.85cm, minimum height=0.50cm,
      font=\scriptsize] (qi) at (4.4,2.55) {$q_i$};
\node[draw=red!50, fill=red!8, rounded corners=3pt,
      minimum width=0.85cm, minimum height=0.50cm,
      font=\scriptsize] (kj) at (4.4,1.05) {$k_j$};
\node[draw=blue!65, fill=blue!13, rounded corners=3pt,
      minimum width=1.55cm, minimum height=0.50cm,
      font=\scriptsize] (Rqi) at (6.3,2.55) {$R(\phi_i)\,q_i$};
\node[draw=red!65, fill=red!13, rounded corners=3pt,
      minimum width=1.55cm, minimum height=0.50cm,
      font=\scriptsize] (Rkj) at (6.3,1.05) {$R(\phi_j)\,k_j$};
\draw[-{Stealth[length=4pt]}, blue!60, thick] (qi) -- (Rqi);
\node[font=\scriptsize, blue!65!black, above, yshift=2pt] at
  ({(4.4+6.3)/2}, 2.73) {$R_{\omega}(\phi_i)$};
\draw[-{Stealth[length=4pt]}, red!60, thick] (kj) -- (Rkj);
\node[font=\scriptsize, red!65!black, below, yshift=-2pt] at
  ({(4.4+6.3)/2}, 0.90) {$R_{\omega}(\phi_j)$};
\node[draw=black!45, fill=black!5, rounded corners=3pt,
      minimum width=1.1cm, minimum height=0.55cm,
      align=center, font=\scriptsize] (dot) at (7.9,1.80)
  {$\tilde{q}_i^\top\tilde{k}_j$};
\draw[-{Stealth[length=4pt]}, black!40] (Rqi.east) -- (dot.north west);
\draw[-{Stealth[length=4pt]}, black!40] (Rkj.east) -- (dot.south west);
\node[font=\normalsize, black!55] at (9.6,1.80) {$=$};
\node[draw=green!55!black, fill=green!10, rounded corners=3pt,
      minimum width=2.7cm, minimum height=0.55cm,
      align=center, font=\scriptsize] (res) at (10.6,1.80)
  {$q_i^\top R_{\omega}(\Delta\phi_{ij})\,k_j$};
\draw[-{Stealth[length=4pt]}, black!35] (dot) -- (res);
\node[font=\scriptsize, text=green!35!black, align=left,
      text width=3.6cm, anchor=west] at (12.1,1.80)
  {\textit{depends only on $\Delta\phi_{ij}$}\\[2pt]
   $\Rightarrow$ invariant under $\phi_i \to \phi_i{+}\Delta\phi_0$};
\end{tikzpicture}%
}
\caption{Schematic illustration of RoPE azimuthal invariance.
RoPE rotates each query and key by its particle's $\phi$ before the
dot product; the two rotation matrices cancel, leaving
$R_{\omega}(\Delta\phi_{ij})$. Under any global rotation
$\phi_i \to \phi_i + \Delta\phi_0$, every pairwise difference is unchanged
and the classifier output is exactly invariant.}
\label{fig:rope_schematic}
\end{figure}

\paragraph{Message passing neural network (MPNN) and MPNN+RoPE}
The MPNN represents each event as a fully-connected directed graph
with reconstructed objects as nodes. The base MPNN is
permutation-invariant by construction, but remains sensitive to
absolute azimuthal orientation $\phi$. The MPNN+RoPE variant closes
this gap using \emph{Rotary Position Encoding}
(RoPE)~\cite{Su:2021rope}, achieving exact azimuthal invariance by construction. Each message-passing layer updates
node representations by aggregating neighbourhood information:
\begin{equation}
\mathbf{h}_i^{(l+1)} = \mathrm{LN}\!\left(
\mathbf{h}_i^{(l)} + \mathrm{AGG}_{j \in \mathcal{N}(i)}
f_\theta\!\left(\mathbf{h}_i^{(l)},\, \mathbf{h}_j^{(l)}\right)
\right),
\label{eq:mpnn}
\end{equation}
where AGG and $f_\theta$ depend on the aggregation and message-passing type. RoPE encodes
$\phi$ as a rotation of the query and key vectors before the
dot product:
\begin{equation}
  \tilde{q}_i = R_{\omega}(\phi_i)\, q_i,
  \qquad
  \tilde{k}_j = R_{\omega}(\phi_j)\, k_j,
  \label{eq:rope_def}
\end{equation}
where $R_{\omega}(\phi)$ is a block-diagonal rotation matrix. The
attention score then satisfies
\begin{equation}
  \tilde{q}_i^{\top} \tilde{k}_j
  \;=\;
  q_i^{\top} R_{\omega}(\phi_j - \phi_i)\, k_j,
  \label{eq:rope_score}
\end{equation}
which depends only on $\Delta\phi_{ij} = \phi_j - \phi_i$. Under a
global rotation $\phi_i \to \phi_i + \Delta\phi_0$ every pairwise
difference is unchanged, so the classifier output is exactly
invariant. A single-frequency variant ($\omega = 1$) is used,
appropriate for $\phi \in [0, 2\pi)$ which has a single natural
scale. Absolute $\phi$ is additionally removed from node features
before embedding, closing the remaining path through which orientation
dependence could enter. The mechanism is illustrated in
\cref{fig:rope_schematic}.

\paragraph{Particle Transformer (ParT)}
The Particle Transformer (ParT) is a full transformer encoder in
which every attention layer receives a persistent pairwise physics
bias. The attention logit for head $h$ and particle pair $(i,j)$ is:
\begin{equation}
a_{ij}^{(h)} = \frac{q_i^{(h)\top} k_j^{(h)}}{\sqrt{d_h}}
+ \mathrm{PairEmbed}(e_{ij})^{(h)},
\label{eq:part_attn}
\end{equation}
where $d_h = d_\mathrm{model}/H$ is the per-head dimension,
and $e_{ij}$ is the level-dependent pairwise physics feature vector
(\cref{eq:edge_core,eq:edge_det}), linearly projected to a
per-head scalar bias through a shared network. The
bias is active at every head and layer, providing a persistent
pairwise geometric inductive bias that the vanilla attention mechanism
lacks.

ParT achieves exact azimuthal invariance through two complementary
mechanisms. First, all pairwise features entering $e_{ij}$ are
coordinate differences or Lorentz invariants, and are therefore
rotation-invariant by construction: the pair bias never sees
absolute $\phi$. Second, only rotationally invariant quantities are
retained as node encoder inputs: absolute orientation-carrying
components are stripped before embedding,
eliminating the only remaining path through which orientation
dependence could enter. The combination guarantees exact invariance
algebraically, without data augmentation or approximate methods.

Graph-level pooling uses class attention~\cite{Zhai:2022} via a
learnable summary token attending to all particle tokens through
dedicated cross-attention blocks. 
The global event features are projected to the hidden dimension and added
to particle token embeddings before the first attention layer.
\paragraph{LorentzNet}
LorentzNet is the second architecture with explicit physics priors, enforcing full Lorentz group invariance, the strongest symmetry constraint of the six models. It treats reconstructed objects as nodes of a fully-connected graph and processes them through a stack of Lorentz Group Equivariant Blocks (LGEBs), each performing a coupled update of a scalar node representation $h_i$ and the 4-momentum coordinate $x_i$:
\begin{align}
  m^l_{ij} &= \varphi_e\!\left(
              h^l_i,\, h^l_j,\,
              \psi\!\left(\|x^l_i-x^l_j\|^2_M\right),\,
              \psi\!\left(\langle x^l_i,x^l_j\rangle_M\right)
            \right), \label{eq:lgeb_msg} \\
x^{l+1}_i &\leftarrow x^l_i + c\cdot\sum_{j}
            \varphi_x(m^l_{ij})\cdot (x^l_i - x^l_j), \\
  h^{l+1}_i &\leftarrow h^l_i + \varphi_h\!\Big(
              h^l_i,\;
              \sum_j \varphi_m(m^l_{ij})\cdot m^l_{ij},\;
              s_i
            \Big), \label{eq:lgeb_node}
\end{align}
where $\|\cdot\|^2_M$ and $\langle\cdot,\cdot\rangle_M$ denote the
Minkowski norm squared and inner product respectively, $\psi(x) =
\mathrm{sgn}(x)\log(|x|+1)$ maps broad Lorentz-invariant
distributions to a numerically stable range, and $s_i$ are the
original invariant node scalars passed at every layer. The edge
messages are built exclusively from Minkowski invariants, so $h_i$
is Lorentz-invariant and $x_i$ transforms as a 4-vector by
construction. The pooled event representation is therefore exactly
Lorentz-invariant, subsuming azimuthal invariance as a special case.

The node encoder maps the invariant scalars to the hidden dimension; kinematic
quantities that break Lorentz invariance are
excluded from node attributes. Graph-level pooling
uses global mean aggregation over the final node representations.

It is a natural question whether enforcing full Lorentz group invariance is the appropriate choice here, and whether it provides additional benefits over the weaker azimuthal constraint enforced by ParT and MPNN+RoPE.
As discussed above, boost invariance is already
broken in reconstructed events, so the stronger prior may discard
genuinely discriminating information. This benchmark directly
addresses this question by comparing all three levels of symmetry
constraint under controlled conditions.

\section{Event representation}\label{sec:representation}

The event content described in \cref{sec:event_content} is
organised differently for each of the six architectures introduced
above (\cref{sec:arch_sym}), depending on whether the model
operates on a flat concatenated vector or a set of graph nodes.

\paragraph{Flat representations}
The \tth\ multilepton final state contains particles with no physically preferred ordering, so any sensitivity of the classifier to the arbitrary input sequence constitutes spurious learned structure. \emph{Flat models} (MLP, XGBoost) receive a single concatenated feature vector formed by stacking all 14 particle slots in a fixed order: four leptons, two taus, and eight jets, each group sorted by decreasing $p_T$. Within each particle type the $p_T$ ordering carries physical meaning (the leading lepton has special kinematic status, the leading jet is most likely to be from a top decay), and flat models can exploit this implicitly. Sorting by $p_T$ does render the input invariant to the arbitrary ordering of the raw event record, but this invariance is achieved only through the sort itself: it becomes unstable whenever two particles have comparable $p_T$, since a small detector-resolution fluctuation can swap their slot assignment and produce a discontinuous jump in the feature vector for a near-identical event. The fixed slot layout also provides no explicit mechanism to compute inter-particle distances or pairwise quantities.

\paragraph{Graph representations}

\emph{Graph models} (MPNN, ParT, LorentzNet) receive the same raw
event object features organized as a set of node feature vectors, one
per present object, connected by a directed fully-connected edge
structure. Graph models are permutation-invariant by construction:
attention and message passing are order-agnostic operations, and no
positional slot index is assigned to individual objects. All graph
models additionally use \emph{object type embeddings} to distinguish
particle species without breaking permutation invariance. Object type
identity is kept entirely outside the feature vector: rather than
appending a one-hot type indicator to the per-node kinematic features,
each object's type is looked up in a learnt embedding table
$E\in\mathbb{R}^{5\times d}$ (one row per type: \texttt{electron},
\texttt{muon}, \texttt{tau}, \texttt{jet}, \texttt{padding}, with the
padding row fixed to zero) and added directly to the hidden
representation after node encoding,
\begin{equation}
\mathbf{h}_i = \mathrm{Linear}(\mathbf{x}_i) + E_{\,\mathrm{type}(i)},
\label{eq:type_embedding}
\end{equation}
where $\mathbf{x}_i$ is the raw kinematic feature vector of object
$i$ and $\mathrm{type}(i)\in\{0,\dots,4\}$ its object-type index. Since
$E_{\,\mathrm{type}(i)}$ depends only on object identity and not on
the object's position within the input ordering, this construction
supplies type information without introducing a slot- or
order-dependent signal, preserving exact permutation invariance. The
MPNN uses no pairwise feature mechanism, unlike ParT: node
representations are updated purely through message passing, with
global features (where available) concatenated only at the pooling
stage. ParT recomputes its own set of pairwise physics observables
on-the-fly via a dedicated \texttt{PairwisePhysicsFeatures} module
that has access to the full node feature matrix, producing the
pairwise pair bias described in \cref{eq:edge_core,eq:edge_det}.
LorentzNet has the most restricted representation of the three: node
attributes are limited to Lorentz-invariant scalars, with no pairwise
mechanism and no global features at any level, while the
four-momenta themselves are propagated separately as equivariant
coordinates rather than as node attributes (\cref{sec:feature_hierarchy}).
Absent particle slots are masked and excluded from all computations.
The feature content available to each architecture at each hierarchy
level is described in the following section.

\section{Feature Set Hierarchy}\label{sec:feature_hierarchy}

To enable fair comparison across architectures, we define a four-level
feature hierarchy that provides identical raw information to all model
types at each level; what each architecture computes from it is an
architectural choice. The hierarchy is illustrated in \cref{fig:feature_hierarchy}.
Object type identity (electron, muon, $\tau_\mathrm{had}$, jet) is
supplied separately from this hierarchy at every level, via the object
type embedding described in \cref{sec:arch_sym}, and is therefore
omitted from \cref{fig:feature_hierarchy} and the level-by-level
descriptions below.

\begin{figure*}[t]
\centering
\resizebox{\textwidth}{!}{%
\begin{tikzpicture}[font=\small]
  \colorlet{cCore}{blue!15}
  \colorlet{cObj} {green!20}
  \colorlet{cKin} {orange!20}
  \colorlet{cFull}{violet!15}
  \colorlet{cPair}{teal!18}
  \fill[cCore] (0,4.05) rectangle (3.2,4.65);
  \draw[black!25] (0,4.05) rectangle (3.2,4.65);
  \node[anchor=center,font=\scriptsize] at (1.6,4.35) {$E,\;p_x,\;p_y,\;p_z$};
  \node[anchor=east,font=\ttfamily\small] at (-0.15,4.35) {\fs{core}};
  \fill[cCore] (0,3.35)   rectangle (3.2,3.95);
  \fill[cObj]  (3.2,3.35) rectangle (5.4,3.95);
  \draw[black!25] (0,3.35) rectangle (5.4,3.95);
  \draw[black!25] (3.2,3.35) -- (3.2,3.95);
  \node[anchor=center,font=\scriptsize] at (1.6,3.65) {$E,\;p_x,\;p_y,\;p_z$};
  \node[anchor=center,font=\scriptsize] at (4.3,3.65) {charge,~$b$-tag};
  \node[anchor=east,font=\ttfamily\small] at (-0.15,3.65) {\fs{object}};
  \fill[cCore] (0,2.65) rectangle (3.2,3.25);
  \fill[cObj]  (3.2,2.65) rectangle (5.4,3.25);
  \fill[cKin]  (5.4,2.65) rectangle (8.8,3.25);
  \draw[black!25] (0,2.65) rectangle (8.8,3.25);
  \draw[black!25] (3.2,2.65) -- (3.2,3.25);
  \draw[black!25] (5.4,2.65) -- (5.4,3.25);
  \node[anchor=center,font=\scriptsize] at (1.6,2.95) {$E,\;p_x,\;p_y,\;p_z$};
  \node[anchor=center,font=\scriptsize] at (4.3,2.95) {charge,~$b$-tag};
  \node[anchor=center,font=\scriptsize] at (7.1,2.95) {$p_T,\;\eta,\;\phi,\;m$};
  \node[anchor=east,font=\ttfamily\small] at (-0.15,2.95) {\fs{detector}};
  \fill[cCore] (0,1.75) rectangle (3.2,2.55);
  \fill[cObj]  (3.2,1.75) rectangle (5.4,2.55);
  \fill[cKin]  (5.4,1.75) rectangle (8.8,2.55);
  \fill[cFull] (8.8,1.75) rectangle (12.2,2.55);
  \draw[black!25] (0,1.75) rectangle (12.2,2.55);
  \draw[black!25] (3.2,1.75) -- (3.2,2.55);
  \draw[black!25] (5.4,1.75) -- (5.4,2.55);
  \draw[black!25] (8.8,1.75) -- (8.8,2.55);
  \node[anchor=center,font=\scriptsize] at (1.6,2.15) {$E,\;p_x,\;p_y,\;p_z$};
  \node[anchor=center,font=\scriptsize] at (4.3,2.15) {charge,~$b$-tag};
  \node[anchor=center,font=\scriptsize] at (7.1,2.15) {$p_T,\;\eta,\;\phi,\;m$};
  \node[anchor=center,font=\scriptsize,align=center] at (10.5,2.15)
    {MET,~$n_\text{jets}$,~$n_b$\\$n_\ell$,~$n_\tau$,~$HT$};
  \node[anchor=east,font=\ttfamily\small] at (-0.15,2.15) {\fs{full}};
  \draw[black!30, thin] (-1.5,1.55) -- (12.2,1.55);
  \node[anchor=west,font=\scriptsize\itshape,text=black!55] at (0,1.28)
    {Model-specific feature access:};
  \fill[cCore] (0,0.20)   rectangle (3.2,1.00);
  \fill[black!8] (3.2,0.20) rectangle (5.4,1.00);
  \fill[black!8] (5.4,0.20) rectangle (8.8,1.00);
  \fill[black!8] (8.8,0.20) rectangle (12.2,1.00);
  \draw[black!25] (0,0.20) rectangle (12.2,1.00);
  \draw[black!25] (3.2,0.20) -- (3.2,1.00);
  \draw[black!25] (5.4,0.20) -- (5.4,1.00);
  \draw[black!25] (8.8,0.20) -- (8.8,1.00);
  \node[anchor=center,font=\scriptsize,align=center] at (1.6,0.60)
    {$+\,\phi$ \\$-\,p_x,\,p_y$};
  \node[anchor=center,font=\scriptsize,text=black!38,align=center] at (4.3,0.60)
    {No \\ changes};
  \node[anchor=center,font=\scriptsize,text=black!38,align=center] at (7.1,0.60)
    {No \\ changes};
  \node[anchor=center,font=\scriptsize,text=black!38,align=center] at (10.5,0.60)
    {No \\ changes};
  \node[anchor=east,font=\small\bfseries,align=right,text width=1.7cm] at (-0.15,0.60) {MPNN\\+RoPE};
  \draw[black!30, thin] (-1.5,0.10) -- (12.2,0.10);
  \fill[cCore]    (0,-1.50) rectangle (3.2,0.10);
  \fill[cObj] (3.2,-1.50) rectangle (5.4,0.10);
  \fill[cKin]  (5.4,-1.50) rectangle (8.8,0.10);
  \fill[black!8] (8.8,-1.50) rectangle (12.2,0.10);
  \draw[black!25] (0,-1.50) rectangle (12.2,0.10);
  \draw[black!25] (3.2,-1.50) -- (3.2,0.10);
  \draw[black!25] (5.4,-1.50) -- (5.4,0.10);
  \draw[black!25] (8.8,-1.50) -- (8.8,0.10);
  \node[anchor=center,font=\scriptsize,align=center] at (1.6,-0.70)
    {$+\Delta R,\log k_T,$\\$\;\log z,\;\log m^2$\\ - $p_x$, $p_y$};
  \node[anchor=center,font=\scriptsize,align=center] at (4.3,-0.70)
    {$+\;q_i{\cdot}q_j$\\(5 pairs total)};
  \node[anchor=center,font=\scriptsize,align=center] at (7.1,-0.70)
    {$+\Delta\eta,\;\Delta\phi,\;- \phi$\\ (7 pairs total) \\ (detector dim = 3)};
  \node[anchor=center,font=\scriptsize,text=black!38,align=center] at (10.5,-0.70)
    {No changes};
  \node[anchor=east,font=\small\bfseries,align=right,text width=1.7cm] at (-0.15,-0.70) {ParT};
  \draw[black!30, thin] (-1.5,-1.60) -- (12.2,-1.60);
  \fill[blue!12]  (0,-2.40) rectangle (3.2,-1.60);
  \fill[black!8] (3.2,-2.40) rectangle (5.4,-1.60);
  \fill[cKin]  (5.4,-2.40) rectangle (8.8,-1.60);
  \fill[violet!15](8.8,-2.40) rectangle (12.2,-1.60);
  \draw[black!25] (0,-2.40) rectangle (12.2,-1.60);
  \draw[black!25] (3.2,-2.40) -- (3.2,-1.60);
  \draw[black!25] (5.4,-2.40) -- (5.4,-1.60);
  \draw[black!25] (8.8,-2.40) -- (8.8,-1.60);
  \node[anchor=center,font=\scriptsize,align=center] at (1.6,-2.00) {+ $\psi(m^2)$ (scalar)};
  \node[anchor=center,font=\scriptsize,text=black!38,align=center] at (4.3,-2.00)
    {No \\ changes};
  \node[anchor=center,font=\scriptsize,align=center] at (7.1,-2.00)
    {---~All excluded~---\\not Lorentz-inv.};
  \node[anchor=center,font=\scriptsize,align=center] at (10.5,-2.00)
    {---~All excluded~---\\not Lorentz-inv.};
  \node[anchor=east,font=\small\bfseries] at (-0.15,-2.00) {LN};
\end{tikzpicture}%
}
\caption{Feature set hierarchy. Each row accumulates the features of all
rows above it; block colours indicate the level that introduced each group.
\textbf{MPNN+RoPE}: absolute $\phi$ is computed from $p_x,p_y$ and used for
rotary position encoding, while $p_x,p_y$ themselves are not added in
node encoder inputs at every level, closing the node encoder path to
absolute azimuthal orientation. \textbf{ParT pair bias}: pairwise
features computed on-the-fly at each level, providing the additive attention
bias (\cref{eq:edge_core,eq:edge_det}); the bias grows from
4 features at \fs{core} (matching the original ParT~\cite{Qu:2022mxj}),
to 5 at \fs{object} with the addition of $q_i{\cdot}q_j$, and to 7 at
\fs{detector} where $\Delta R$ is decomposed into $\Delta\eta$ and
$\Delta\phi$. Node stripping ($-p_x$, $-p_y$ at all levels; $-\phi$ from
\fs{detector} onward) closes the node encoder path to absolute azimuthal
orientation, ensuring exact azimuthal invariance. \textbf{LN}: LorentzNet
uses 4-momenta as equivariant coordinates and only the Lorentz-invariant
scalars as node attributes; \fs{detector} and \fs{full} are therefore
excluded.}
\label{fig:feature_hierarchy}
\end{figure*}

\paragraph{\fs{core} --- raw four-momenta.}
Each object slot carries the Cartesian four-momentum
$(E,\,p_x,\,p_y,\,p_z)$ (\cref{sec:event_content}), giving flat models
56 columns and graph models a node dimension of 4. All graph models
additionally have access to the object type embedding
(\cref{sec:representation}) already at this level. MPNN+RoPE computes
$\phi$ from the Cartesian components for use in the rotary position
encoding of \cref{eq:rope_def,eq:rope_score}, and doesn't include
$p_x$, $p_y$ in the node encoder inputs to preserve exact azimuthal
invariance. ParT internally derives four pairwise features from the
four-momenta:
\begin{equation}
e_{ij} = [\,\Delta R_{ij},\;\log k_{T,ij},\;\log z_{ij},\;\log|m^2_{ij}|],
\label{eq:edge_core}
\end{equation}
and likewise strips $p_x$, $p_y$ from the node encoder inputs to
preserve exact azimuthal invariance. LorentzNet uses only the compressed
invariant mass-squared $\psi(m^2)$ as its node scalar (scalar
dimension~1); the compression function $\psi$ is defined in
\cref{sec:arch_sym}.
The $p_x$, $p_y$, $p_z$, $E$
are passed as Lorentz-equivariant coordinates and are not part of the
invariant node attribute vector in LorentzNet.

\paragraph{\fs{object} --- $\&$ charge/flavor.}
The per-object charge or $b$-tag score (\cref{sec:event_content}) is
appended, bringing the per-object count to 5. Flat models receive 70
columns; graph models have a node dimension of 5. ParT gains access
to the charge product $q_i{\cdot}q_j$ as a fifth pairwise feature,
giving a 5-dimensional pair bias. LorentzNet adds charge or $b$-tag
score as a second invariant scalar alongside $\psi(m^2)$ (scalar
dimension~2).

\paragraph{\fs{detector} --- $\&$ detector coordinates.}
The detector coordinates $(p_T,\,\eta,\,\phi,\,m)$
(\cref{sec:event_content}) are added per object as shared input
features, bringing the per-object count to 9 (flat: 126 columns; node
dimension: 9). With $\eta$ and $\phi$ now available, $\Delta\eta_{ij}$
and $\Delta\phi_{ij}$ are added to the ParT pair bias alongside the
existing $\Delta R_{ij}$, carrying distinct physical meaning at the
event level beyond the combined
$\Delta R_{ij}=\sqrt{\Delta\eta_{ij}^2+\Delta\phi_{ij}^2}$. The full
ParT pair bias becomes:
\begin{equation}
e_{ij} = [\,\Delta\eta_{ij},\;\Delta\phi_{ij},\;\Delta R_{ij},\;
\log k_{T,ij},\;\log z_{ij},\;\log|m^2_{ij}|,\;q_i{\cdot}q_j],
\label{eq:edge_det}
\end{equation}
giving a 7-dimensional pair bias used as additive attention biases at
every head and layer. Of these shared features, only $\phi$ is
not added to the ParT node encoder inputs, to preserve exact
azimuthal invariance; $p_T$, $\eta$, and $m$ are retained as node
attributes. The detector coordinates are not invariant under the full
Lorentz group and are therefore excluded from the LorentzNet node
attribute vector; its scalar dimension remains~2.

\paragraph{\fs{full} --- $\&$ global event features.}
The six global event-level quantities defined in \cref{sec:event_content} are exposed as input features for the first time at this level. For flat models and MPNN variants, all six are passed as global scalars (flat: 132 columns total; graph global dimension: 6). ParT projects the global features to the hidden dimension and adds them to particle token embeddings before the first attention layer. As at \fs{detector}, LorentzNet is not evaluated at this level: none of the six global features are Lorentz-invariant even in principle ($E_{\mathrm{T}}^{\mathrm{miss}}$ and $\rm HT$ depend on the lab-frame transverse plane, and the multiplicity counts $n_\mathrm{jets}$, $n_\mathrm{bjets}$, $n_\mathrm{leptons}$, $n_\mathrm{taus}$ depend on detector acceptance). In preliminary tests, passing these non-invariant global features to LorentzNet degraded its performance relative to the \fs{object}-level configuration, consistent with the node-level Lorentz-invariance constraint being genuinely load-bearing rather than an incidental restriction.
\section{Hyperparameter optimisation and training}\label{sec:hpo}

The dataset is split 80\%/10\%/10\% into training, validation, and
test sets by stratified sampling across (channel, process) cells
(\cref{sec:dataset}). The training split is used to fit model
parameters; the validation split is used for early stopping, HPO
pruning, and as the HPO objective (validation ROC AUC), which
correlates well with profile likelihood rankings though not always
perfectly, as \cref{sec:results} shows; the test split is held out
entirely from both training and HPO, and is used only once, at the
end, for final evaluation.
All architectures are independently optimised using Optuna
\cite{Akiba:2019optuna} with the TPE sampler in a shared SQLite
database, with validation ROC AUC as the objective. The feature-level
reuse of hyperparameters across the four levels of the hierarchy is
detailed in \cref{sec:feature}; the data-fraction reuse
scheme used in the scaling study is described in \cref{sec:scaling}.
The full search spaces are given in \cref{app:hpo}.
Each final model configuration is trained five times with independent
dataset splits, using the same splitting procedure described above;
all results are reported as the mean and standard deviation over the
five seeds (\cref{sec:training}).

\paragraph{XGBoost}
The configurable parameters are tree depth, learning rate, row and
column subsampling fractions, and regularisation strengths
($\alpha$, $\lambda$). Early stopping monitors validation log-loss
over up to 5{,}000 trees.
\paragraph{MLP}
Up to six hidden layers are stacked, each applying a linear
transformation followed by an activation function, optional LayerNorm,
and dropout. Residual connections are added where widths match, with a
linear projection otherwise. The number of layers, hidden width,
dropout rate, activation (GELU, ReLU, SiLU, or ELU), LayerNorm, and
residual connections are all searched via HPO, alongside the learning
rate, weight decay, and batch size.
\paragraph{MPNN and MPNN+RoPE}
The HPO search covers convolution type (\textit{GCN}~\cite{Kipf:2017}, \textit{GAT}~\cite{Velickovic:2018}, \textit{GraphSAGE}~\cite{Hamilton:2017}, \textit{TransformerConv}~\cite{Shi:2020masked}), pooling mechanism (mean, max, sum, score-weighted attention, and CLS via a learnable summary token), number of layers, hidden dimension, and activation function. Conditional parameters (number of attention heads, FFN multiplier, and CLS readout layers) are searched only when the relevant conv type or pooling is selected. HPO consistently selects TransformerConv across all feature levels; each layer applies multi-head attention followed by a position-wise FFN sublayer, both with Pre-LN and residual connections. For MPNN+RoPE, the aggregation type is fixed to TransformerConv (RoPE requires the scaled dot-product structure), CLS pooling is excluded (the dense class-token cross-attention is not well-defined under single-frequency RoPE), and the remaining search space mirrors the base MPNN.
\paragraph{ParT}
The HPO search covers hidden dimension, number of attention heads
(constrained such that the head dimension is at least 32), number of
transformer layers, number of class-attention layers, FFN multiplier,
separate dropout rates for attention and FFN sublayers, and the
learning rate, weight decay, batch size, warmup length, and cosine
period. The pair embedding network dimensions are fixed to
$[64, 64, 64]$ to reduce the search space.
\paragraph{LorentzNet}
The HPO search covers hidden dimension, number of LGEB layers, the
coordinate scale hyperparameter $c$, dropout rate, pooling mechanism
(mean, max, sum), and the learning rate, weight decay, batch size,
cosine period, and minimum learning rate fraction. The warmup length
range is wider than for other architectures (0--40 epochs),
reflecting the slower convergence of the Minkowski inner product
computations.

\subsection{Training procedure}\label{sec:training}
Each classifier is trained as a probabilistic predictor
$p(y\mid x)$, where $x$ is the input feature vector and $y \in
\{0,1\}$ denotes the background and signal class respectively.
Training maximises the weighted log-likelihood
\begin{equation}
\max_\theta \sum_{i=1}^{N} w_i \log p(y_i \mid x_i;\, \theta),
\label{eq:loss}
\end{equation}
where $w_i$ is the weight of event $i$~\eqref{eq:weights}, determined by
its parent process $\mathrm{proc}(i)$, assigned so that the weighted
sample reflects the physically expected process rates. This is
equivalent to minimising the weighted binary cross-entropy loss with
normalisation $1/\sum_i w_i$.

\section{Statistical Evaluation}\label{sec:statistics}
The primary metric of interest is the statistical uncertainty $\sigma(\mu)$~\cite{Cowan:2010js} expected for a given luminosity $\mathscr{L}$ on
the signal strength $\mu = \sigma_{t\bar{t}H}/\sigma_{t\bar{t}H}^{\mathrm{SM}}$
from a simultaneous six-channel profile likelihood fit of the classifier score performed with
\TREx v1.7.0~\cite{TRExFitter}. All fits share a common \TREx
configuration: the fit type is set to \texttt{SPLUSB} (signal-plus-background fit), the likelihood
is minimised with \texttt{MINUIT}, and the \ttw\ and \ttz\
normalisations are set to the Standard Model values. The classifier output score is discretised using
the TRExFitter \texttt{TransfoD} algorithm with parameters $z_b = z_s = 10$
\cite{TRExFitter}, which iteratively merges adjacent fine bins from the background-rich (low-score) end
of the distribution until $Z = z_b\, n_b/N_b + z_s\, n_s/N_s$ exceeds unity,
where $n_b$ ($n_s$) and $N_b$ ($N_s$) are the background (signal) yield in
the current bin and the total background (signal) yield, respectively. This effectively caps each merged bin at roughly $1/z_b$ ($1/z_s$) of the
total yield. An identical binning
configuration across all channels and architectures ensures a fair comparison
of classifier shapes.

The binned extended likelihood over the three processes ($t\bar tH$,
$t\bar tW$, $t\bar tZ$) is
\begin{equation}
\mathcal{L}(\mu, \boldsymbol{\theta}) =
\prod_{c=1}^{6}\prod_{k=1}^{K}
\mathrm{Pois}\!\left(n_{ck} \mid \nu_{ck}(\mu,
\boldsymbol{\theta})\right)
\cdot \prod_{c,k} \mathrm{Gam}(\gamma_{ck}),
\label{eq:likelihood}
\end{equation}
with $K$ the number of bins in the classifier score distribution and
nuisance vector $\boldsymbol{\theta} =
\{\gamma_{ck}\}$, where $\gamma_{ck}$
is a single, pooled-over-process Barlow--Beeston lite MC-statistical
nuisance per bin~\cite{Barlow:1993dm,Conway:2011in}, propagated via the
\texttt{autoMCStats}~\cite{Cranmer:2012sba} approach. Performance is evaluated on the Asimov
dataset for the expected event yields at luminosity $\mathscr{L}$ with $\mu = 1$ (the MC simulation ground truth). This is an established procedure for measuring the uncertainty associated with estimating $\mu$.
\cref{app:stat-model} gives a brief overview of the details.

Results are reported as $\varrho_\mu = \sigma(\mu)/\sigma_\mathrm{ref}$, where
$\sigma(\mu)$ is the uncertainty from the binned classifier likelihood fit
and $\sigma_\mathrm{ref}$ is the reference uncertainty from a single
compressed bin per channel (six bins total). With backgrounds fixed to the Standard Model expectations,
$\mu$ can be determined from a single bin per channel, thus combining six
channels analytically only requires that they be treated as
statistically independent, which holds here since no systematic is
shared across channels. The six single-bin sub-likelihoods therefore
combine by simple inverse-variance weighting,
\begin{equation}
\frac{1}{\sigma_\mathrm{ref}^2} = \sum_{c=1}^{6} \frac{1}{(\sigma_\mathrm{ref}^{c})^2},
\label{eq:sigma_ref_combined}
\end{equation}
with $\sigma_\mathrm{ref}^{c}$ given in closed form in
\cref{app:stat_scaling}.\footnote{Cross-validation against a direct
\TREx\ numerical fit of the six-channel, one-bin-per-channel
configuration gives excellent agreement. The
numerical fit itself showed instability at very large luminosities, which the closed-form calculation avoids
by construction.} The ratio
isolates the information gain from classifier shape on top of the channel
categorisation by normalising out the absolute yield scale; a value below
unity indicates improvement over a channel-binned counting experiment.
Reporting $\varrho_\mu$ is convenient for the comparison of ML models, as
the ranking is stable as a function of luminosities (\cref{fig:lumi_scan}),
i.e., comparable across studies.

For the individual channel experiment (\cref{sec:channels}),
each channel is instead evaluated on its own, and the corresponding
$\sigma_\mathrm{ref}^{c}$ is obtained from a single bin containing only
the events from that channel, with $k_{t\bar{t}W}=k_{t\bar{t}Z}=1$ as used throughout the whole analysis. The value of $\sigma_\mathrm{ref}^{c}$ is therefore also computed in closed form.

In the $2\ell$SS$1\tau$, $1\ell2\tau$, and $2\ell$$2\tau$ channels,
the dominant background in a full analysis~\cite{ATLAS:2025eua} arises from non-prompt (fake)
leptons.
We assume that our simulation describes the non-prompt leptons appropriately, being aware that in a full analysis, the non-prompt rate is determined with a data-driven method.

\FloatBarrier
\section{Results}\label{sec:results}

Three main experiments are reported. \cref{sec:feature} varies
the feature set across four levels while holding the training data
fixed at 100\%. \cref{sec:scaling} varies the training data
fraction at the highest feature level.\footnote{For LorentzNet, the
\fs{object} feature level is used instead of the \fs{full} level, as its architecture cannot
accommodate the global event-level features introduced at the
\fs{detector} and \fs{full} levels, and \fs{object} is where it
achieves peak performance.} \cref{sec:channels} evaluates performance across the six
analysis channels and compares multi-channel training against
dedicated single-channel training. Azimuthal rotation sensitivity is
verified on the test set without retraining; details are given in \cref{app:phi}.
Figures in \cref{sec:feature} and \cref{sec:scaling} show two metrics side by side: ROC AUC and $\varrho_\mu$ at $3000~\mathrm{fb}^{-1}$, as subplots within each figure. The other luminosities are omitted from the main text, as \cref{app:lumi-scan} shows that the relative ranking of
architectures is unchanged across the luminosity range $[10,5000]$~fb$^{-1}$, so a single luminosity is sufficient to
convey the result. All six architectures are shown together, with
unconstrained models (MLP, XGBoost, MPNN) and symmetry-constrained
variants (MPNN+RoPE, ParT, LorentzNet) distinguished by marker style
and colour. Shaded bands and error bars represent the standard
deviation over five independent training runs with different dataset
splits.
\FloatBarrier
\subsection{Feature-set ablation}\label{sec:feature}

Each architecture is trained on 100\% of the available data at each
of the four feature levels (\fs{core}, \fs{object}, \fs{detector},
\fs{full}). Hyperparameters optimised on the \fs{full} feature set
are fixed across the \fs{detector} and \fs{full} levels; the
\fs{core} and \fs{object} levels use hyperparameters selected on the
\fs{core} subset, since the optimal configuration can differ when
only raw four-momenta are available, or global features are included. LorentzNet is evaluated at \fs{core} and \fs{object} only, as
discussed in \cref{sec:feature_hierarchy}.

The experiment isolates the information contribution of each feature
group and reveals how each architecture's inductive bias interacts
with the available input representation: in particular, whether the
pair bias of ParT, the Lorentz-invariant constraint of LorentzNet,
or the azimuthal symmetry of MPNN+RoPE provides consistent advantages
across the hierarchy.

\begin{figure*}[!ht]
\centering
\includegraphics[width=0.49\textwidth]{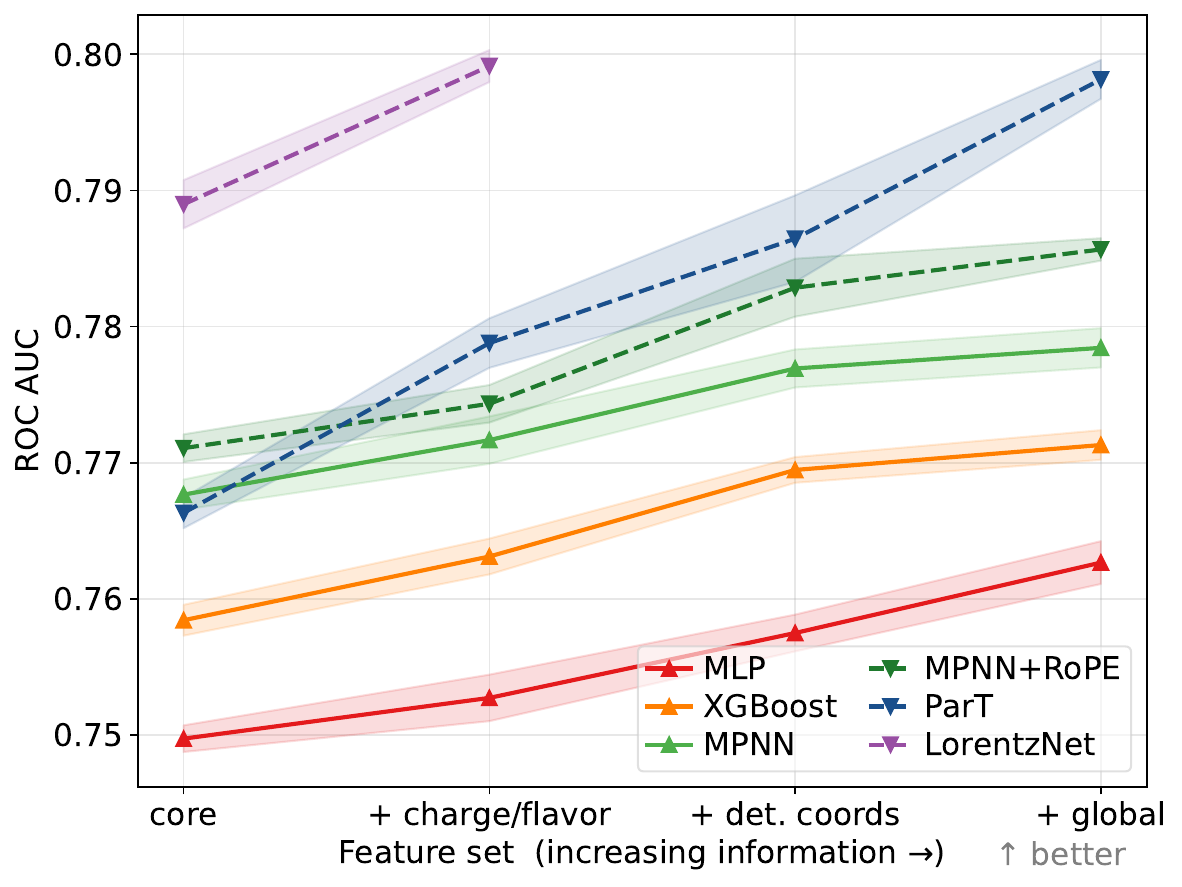}
\includegraphics[width=0.49\textwidth]{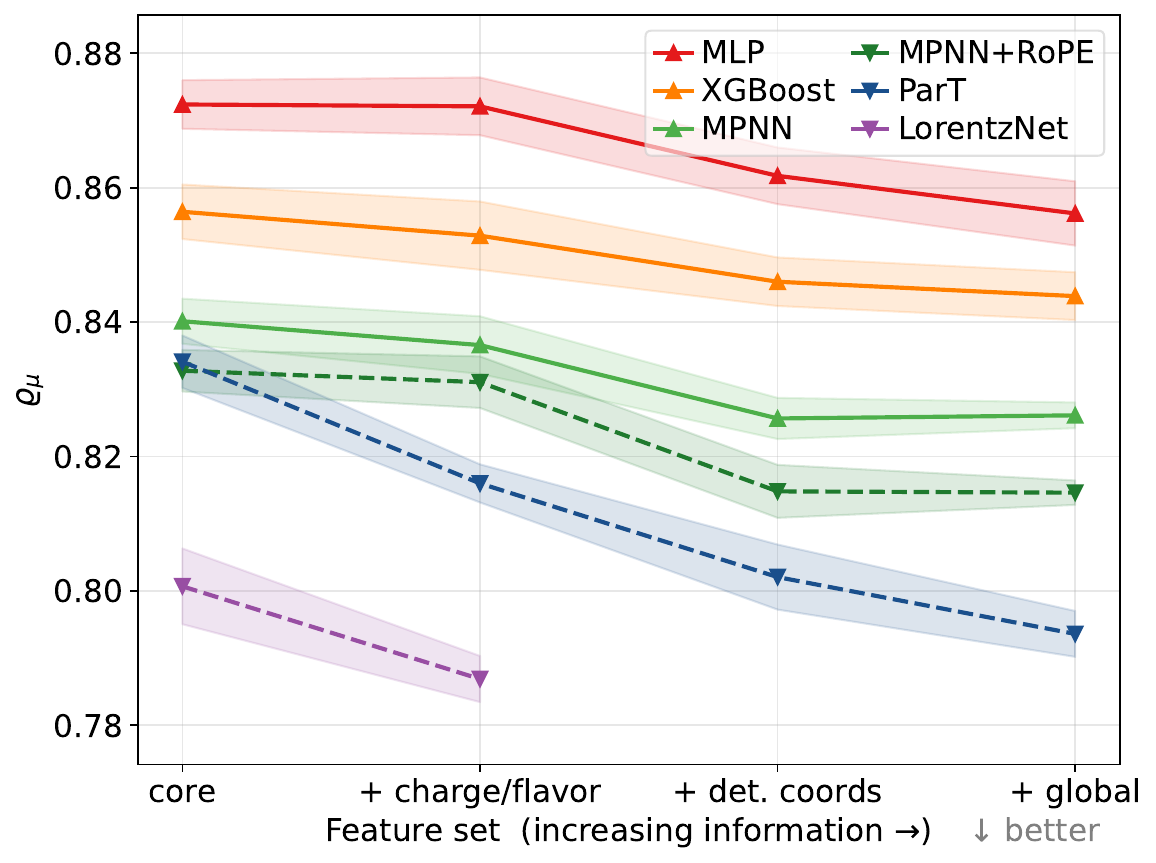}
\caption{ROC AUC (left), and $\varrho_\mu$ at $\mathscr{L}=3000~\mathrm{fb}^{-1}$
(right, HL-LHC projection) versus feature set. All six architectures
are shown together; unconstrained models (MLP, XGBoost, MPNN) and
symmetry-constrained variants (MPNN+RoPE, ParT, LorentzNet). Shaded bands show the
standard deviation over five dataset splits. LorentzNet is not
evaluated at the \fs{detector} and \fs{full} levels as those features
are not Lorentz-invariant.}
\label{fig:feature_auc}
\end{figure*}
\paragraph{Unconstrained architectures.}
All three unconstrained architectures improve monotonically across
the feature hierarchy (\cref{fig:feature_auc}). XGBoost leads MLP at
every level by a consistent margin. MPNN outperforms XGBoost at every
feature level, with a comparable rate of improvement across the
hierarchy, demonstrating that graph-based message passing extracts
more information from the same raw features even without symmetry
constraints.

\paragraph{Constrained architectures.}
MPNN+RoPE outperforms the unconstrained MPNN at every feature level,
by a margin that widens toward the later levels: RoPE's azimuthal
invariance encoding provides a consistent gain over the unconstrained
baseline that grows as more detector and global information becomes
available, indicating that enforcing this symmetry remains beneficial
even as the raw feature content increases.

At \fs{core}, MPNN+RoPE marginally outperforms ParT: with only raw four-momenta available, the azimuthal symmetry encoding provides a small advantage, while ParT's default pair bias (built from Lorentz-invariant pairwise features, since $p_x$, $p_y$ are already stripped from its node encoder at this level; \cref{sec:feature_hierarchy}) offers limited additional discrimination over node-only message passing. This
ordering reverses at \fs{object}, where the charge product
$q_i{\cdot}q_j$ enters the ParT pair bias as a fifth pairwise
feature. Same-sign and opposite-sign lepton pairs are strongly
discriminating in the multilepton final state, and ParT exploits
this directly in the attention bias at every layer, producing a
substantial gain while MPNN+RoPE remains relatively flat. At both
levels, however, LorentzNet leads every other architecture on both
ROC AUC and $\varrho_\mu$, consistent with its Lorentz-invariant
representation being well matched to the available input even before
detector-specific information is introduced.

At \fs{detector}, MPNN+RoPE improves further, roughly in step with
ParT's own gain over the same step, so ParT's lead over MPNN+RoPE
remains essentially unchanged from \fs{object}. Since $\phi$ is
already available at earlier feature levels, MPNN+RoPE's gain is
more likely driven by the additional detector coordinates ($p_T$,
$\eta$, $m$) becoming available as node features, rather than by any
change in $\phi$ availability. The same new node features become
available to ParT at this level, alongside $\Delta\eta$ and
$\Delta\phi$ entering its pair bias, so its comparable gain may stem
from either source.

At \fs{full}, ParT extends its lead over MPNN+RoPE substantially
through stronger global feature integration, projecting global
quantities to the hidden dimension before the first attention layer
rather than concatenating at the pooling stage as MPNN+RoPE does,
though its result still falls short of LorentzNet's \fs{object}-level
performance on $\rho(\mu)$.
LorentzNet is not evaluated beyond \fs{object}, as detector and
global features are excluded by construction for not being fully
Lorentz invariant. Exact results can be found in \cref{app:feature}.

\subsection{Data-scaling study}\label{sec:scaling}

\begin{figure*}[!ht]
\centering
\includegraphics[width=0.49\textwidth]{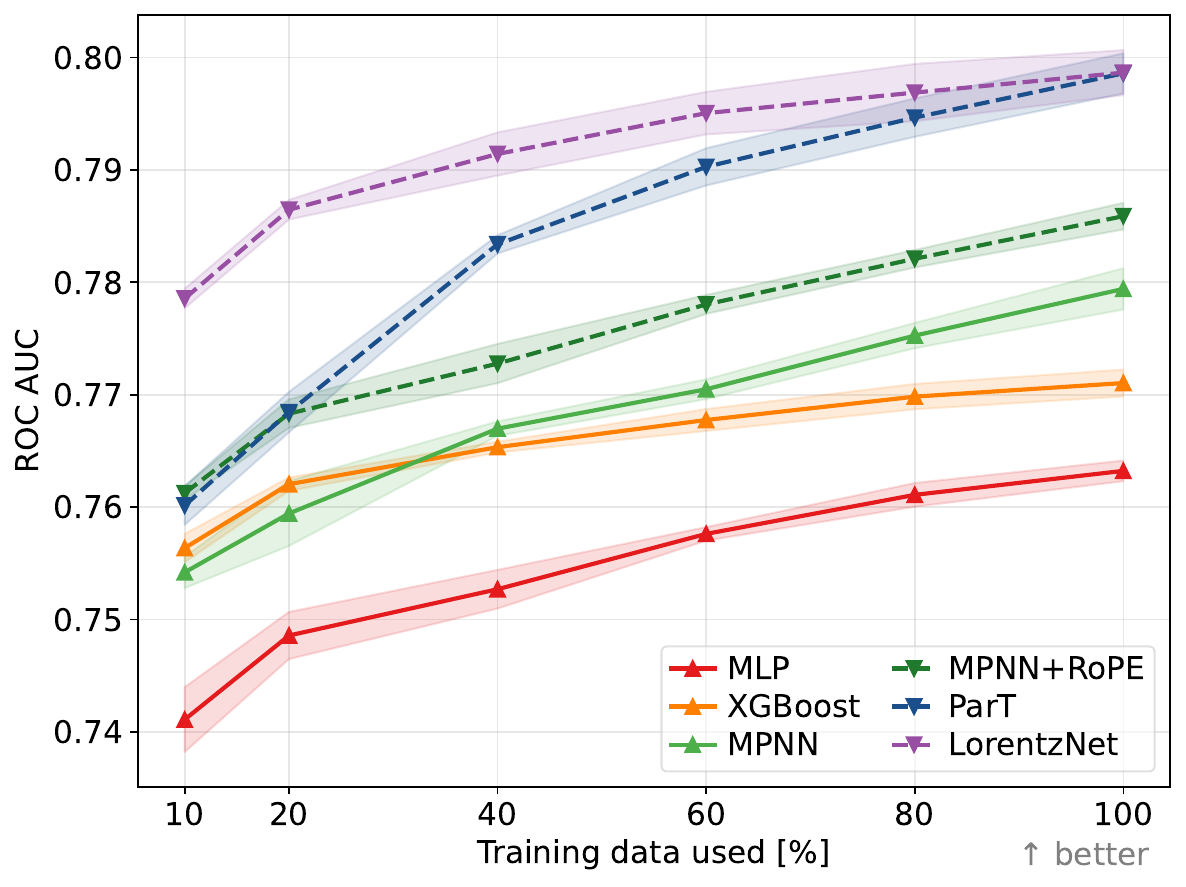}
\includegraphics[width=0.49\textwidth]{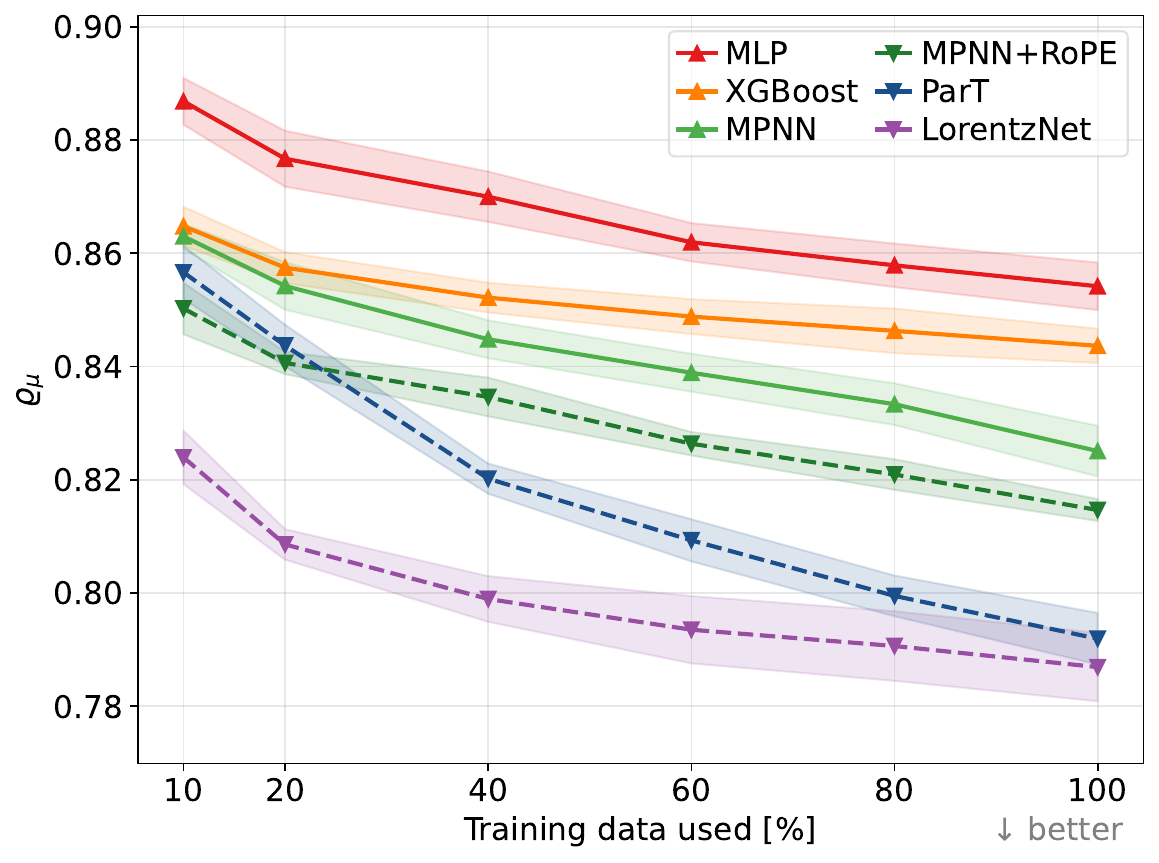}
\caption{ROC AUC (left), and $\varrho_\mu$ at $\mathscr{L}=3000~\mathrm{fb}^{-1}$
(right, HL-LHC projection) versus training data fraction. All six
architectures are shown together; unconstrained models (MLP, XGBoost,
MPNN) and symmetry-constrained variants (MPNN+RoPE, ParT, LorentzNet)
are distinguished by marker style and colour. Shaded bands show the
standard deviation over five dataset splits. LorentzNet is evaluated
at \fs{object}; all other architectures at \fs{full}.}
\label{fig:scaling_auc}
\end{figure*}

All models are trained at the \fs{full} feature level on six fractions
of the available training data: 10\%, 20\%, 40\%, 60\%, 80\%, and
100\%; LorentzNet is trained at \fs{object} as established in \cref{sec:feature}. Hyperparameters are selected from two
independent HPO runs: fractions 10\% and 20\% use hyperparameters
optimised on the 10\% subset, while 40\%, 60\%, 80\%, and 100\% use
hyperparameters from the 100\% subset optimisation. Each fraction is
assigned to the nearer HPO point, since optimal configurations can
differ substantially between the low- and high-data regimes. This
design isolates the effect of sample size from hyperparameter
sensitivity across the scaling curve. The experiment addresses two
questions: how quickly each architecture saturates its achievable
performance, and whether physically motivated inductive biases (pair
bias, RoPE, Lorentz invariance) provide a measurable advantage when
training data are limited. All results are averaged over five
independent seeds; shaded bands in \cref{fig:scaling_auc}
show the corresponding standard deviation.

\paragraph{Unconstrained architectures.}
For all training data fractions, XGBoost visibly outperforms MLP,
consistent with the general strength of gradient-boosted decision
trees on tabular data~\cite{Grinsztajn2022}; MPNN, by contrast, trails
XGBoost only narrowly at 10\%. XGBoost gains less as the training
dataset grows, and is overtaken by MPNN at around 30\% of the
available data. At 100\%, MPNN significantly outperforms both flat
models and shows no sign of saturation, suggesting further gains are
achievable with larger datasets. XGBoost shows the weakest scaling
behaviour among unconstrained architectures in this benchmark,
consistent with the limited depth of representation available to a
fixed-complexity tree ensemble relative to a message-passing model
whose effective capacity grows with the amount of relational
structure it can learn from additional data.

\paragraph{Constrained architectures.}
Remarkably, MPNN+RoPE surpasses XGBoost already at 10\% of the training data (a fraction at which the unconstrained MPNN still trails XGBoost), demonstrating that symmetry constraints alone are sufficient to overcome XGBoost's strength in the low-data regime. ParT and MPNN+RoPE are close together at low fractions, both
meaningfully outperforming MPNN, with the advantage most pronounced
on ROC AUC. As training data increases, ParT closes the gap to
LorentzNet steadily on both metrics: at 100\%, its ROC AUC matches
LorentzNet's, and its $\varrho_\mu$ gap narrows to a small residual,
confirming the trend anticipated in \cref{sec:feature}: ParT's
slower start but steeper scaling allows it to catch up with
sufficient training data, while LorentzNet's advantage, built into
its architecture rather than learned, is largest precisely where
data is scarce. MPNN+RoPE remains below both ParT and LorentzNet at
higher data fractions. No architecture saturates at the available sample
size, suggesting all models would benefit from larger datasets. As
in \cref{sec:feature}, LorentzNet's advantage remains more pronounced
in the profile likelihood metric than in ROC AUC. Exact results can
be found in \cref{app:scale}.

\FloatBarrier
\subsection{Joint versus dedicated training}\label{sec:channels}

A practical question for any classifier analysis is whether training
on a broader, more diverse dataset improves performance on individual
channels, or whether dedicated single-channel training is preferable.
This has direct implications for the feasibility of a foundation
model approach: if cross-channel data sharing is beneficial, a single
model trained on diverse physics processes could generalise
effectively across analysis categories without per-channel
retraining. The six analysis channels differ substantially in signal
purity, dominant background composition, and reconstructed object
multiplicity, and their training sizes span nearly two orders of
magnitude: from 413,740 events in $2\ell\mathrm{SS}0\tau$ down to
3,747 in $2\ell2\tau$ (\cref{fig:channel_counts}).

\begin{figure*}[!ht]
\centering
\includegraphics[width=0.99\textwidth]{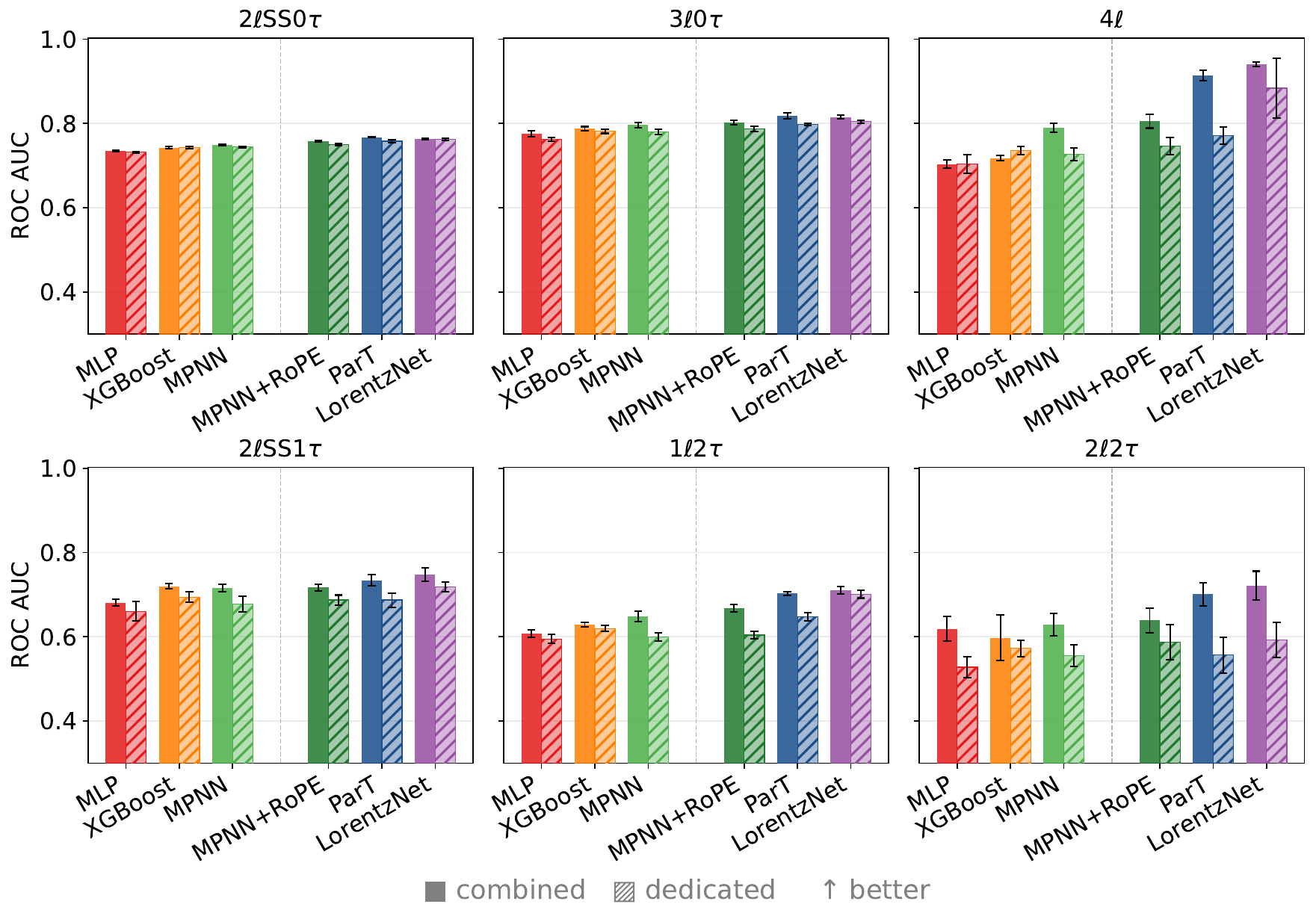}
\caption{Per-channel ROC AUC at the \fs{full} feature level (
\fs{object} for LorentzNet). Double bars correspond to individual
architectures: \emph{left bar} (darker) combined (multi-channel)
training, \emph{right bar} (lighter) dedicated (single-channel)
training. Error bars give the standard deviation over five seeds.}
\label{fig:channel_auc}
\end{figure*}

\begin{figure*}[!ht]
  \includegraphics[width=0.99\textwidth]{%
    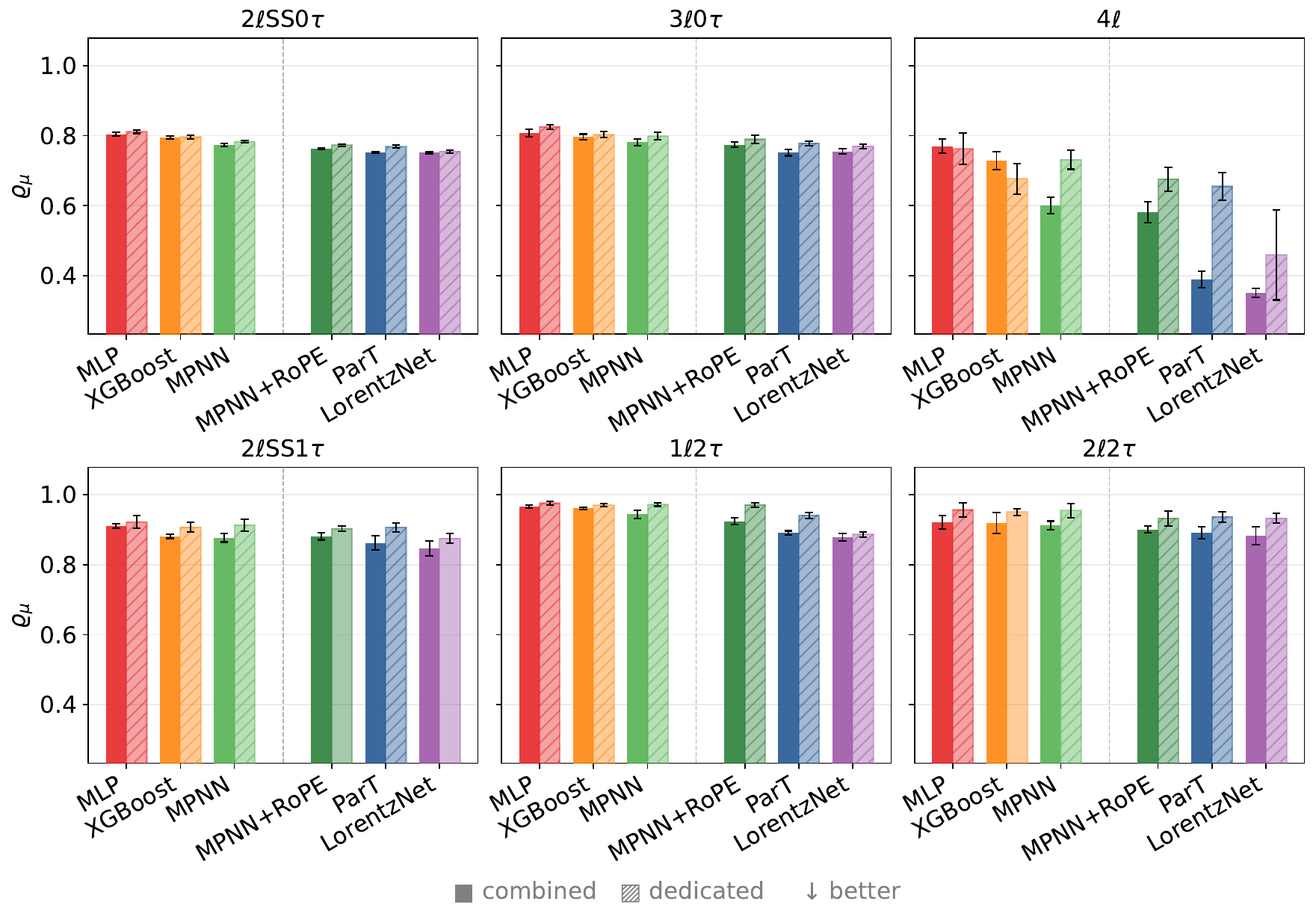}
\caption{Per-channel $\varrho_\mu$ at the
\fs{full} feature level (\fs{object} for LorentzNet) for all six
architectures at $\mathscr{L}=3000~\mathrm{fb}^{-1}$ (HL-LHC projection). \emph{Left bar} (darker): combined (multi-channel)
training. \emph{Right bar} (lighter): dedicated (single-channel)
training. Error bars give the standard deviation over five seeds.
The $4\ell$ channel shows the largest gap between training modes;
dedicated training produces substantially degraded sensitivity for
all architectures due to severe data starvation.}
\label{fig:channel_statunc}
\end{figure*}

\paragraph{Joint training improves per-channel
performance.} \cref{fig:channel_auc,fig:channel_statunc} show
per-channel ROC AUC and profile likelihood sensitivity $\varrho_\mu$
respectively, each as a grid of six panels (one per analysis channel)
with all six architectures compared side by side within each panel,
combined (multi-channel) training shown as solid bars and dedicated
(single-channel) training as hatched bars. Joint training is
beneficial in nearly all cases: the only notable exceptions are MLP
and XGBoost in $4\ell$, where dedicated training holds a small edge
on both, or the results are tied. For the data-rich channels ($2\ell\mathrm{SS}0\tau$,
$3\ell0\tau$), the gap between training modes is modest for all
architectures. For the smallest channels, where dedicated training
suffers from severe data starvation and large seed-to-seed variance,
combined training yields substantial improvements for most
architectures; however, the size of this gain is not uniform, and
instead tracks each architecture's degree of built-in symmetry
constraint. LorentzNet is the clearest case: at $4\ell$, its
dedicated-only result (0.884 AUC) is already the strongest of any
architecture trained on that channel alone, and joint training closes
only a comparatively small remaining gap (to 0.941). This is
consistent with the data-scaling results of \cref{sec:scaling}, where
LorentzNet's Lorentz-equivariant inductive bias lets it extract more
from limited data even without cross-channel sharing. ParT and the
other graph architectures, by contrast, show much larger
combined-versus-dedicated gaps at $4\ell$ (e.g.\ ParT: 0.771 dedicated
to 0.914 combined); here joint training is compensating for a
comparatively weaker inductive bias by supplying more data, rather
than reflecting inherent data efficiency. Full numerical results are
reported in \cref{app:joint}.

\section{Conclusion}\label{sec:conclusion}
We have presented a controlled benchmark framework for event-level
classifier comparison in high-energy physics, using both ROC AUC and
relative profile likelihood uncertainty $\varrho_\mu =
\sigma(\mu)/\sigma_\mathrm{ref}$ as complementary evaluation criterion,
and applied it to \tth\ multilepton signal--background separation at
$\sqrt{s}=14$~TeV across a four-level feature hierarchy. Six
classifiers spanning flat tabular models, generic graph networks, and
architectures with explicit physical symmetry constraints are
evaluated: XGBoost, MLP, MPNN, MPNN+RoPE, ParT, and LorentzNet.
We observed that architecture and feature representation matter substantially. Graph-based and transformer architectures consistently surpass XGBoost as the
default method across all feature levels. At the highest feature level and
$100\%$ training data (\cref{app:summary}), ParT and LorentzNet are
essentially tied for the highest ROC AUC ($0.799$ for both), with LorentzNet
holding only a negligible lead on $\varrho_\mu$ ($0.784$ vs.\ ParT's $0.789$),
indicating good correlation between the two metrics at the top of the
ranking.
The gap between the best and worst architecture amounts to
$\Delta\mathrm{AUC} \approx 0.036$ and $\Delta\varrho_\mu \approx 0.067$,
indicating that architectural choices have a meaningful impact on
statistical sensitivity. 
Enforcing azimuthal invariance via single-frequency
rotary position encoding in an otherwise standard MPNN yields a substantial
gain over unconstrained baselines, suggesting that symmetry constraints are
a significant driver of improvement, separable from architectural
sophistication.
Our next finding is that performance does not saturate with the
available training data. Symmetry-constrained models maintain a
clear advantage at low statistics, confirming that their gains
reflect a genuine inductive bias rather than an artefact of data
abundance.
Finally, we observed that joint training across all six analysis
channels in almost all cases outperforms dedicated per-channel
training, with the largest gains concentrated in data-scarce channels
and graph-based architectures. Together with the data-scaling
results, this suggests potential benefits of training a single shared
model across diverse analysis categories without per-channel
retraining: a step towards the foundation model (in the sense of
machine learning) approach in particle physics.
\backmatter

\bmhead{Acknowledgements}

We acknowledge support by the 
Ministry of Education, Youth and Sports of the Czech
Republic under the project number LM 2023040, and the
Student Grant System SGS25/190/OHK4/3T/35.

\section*{Declarations}

\begin{itemize}
\item \textbf{Funding.} Supported by the Ministry of Education, Youth and Sports of the Czech
Republic under the project number LM 2023040, and the
Student Grant System SGS25/190/OHK4/3T/35.
\item \textbf{Competing interests.} The authors have no competing interests
to declare that are relevant to the content of this article.
\item \textbf{Data availability.} The simulated dataset generated for
this study will be made publicly available on Zenodo under the
CC-BY-4.0 licence upon acceptance of this manuscript, together with
the full simulation configuration files and selection code.
\item \textbf{Author contribution.} L.V.\ conceived the study, developed
and implemented the analysis pipeline, performed all experiments, and wrote
the manuscript. A.S.\ and O.S.\ supervised the project and provided
guidance on physics interpretation and methodology.
\end{itemize}

\bibliography{sn-bibliography}

\begin{appendices}
\crefalias{section}{appendix}
\crefalias{subsection}{appendix}

\section{Dataset and features}\label{app:dataset}
\setcounter{table}{0}
\setcounter{figure}{0}
\setcounter{equation}{0}
\subsection{Per-channel event selection}\label{app:selection}

The \tth multilepton analysis is split into six orthogonal channels
based on lepton and $\tau_{\mathrm{had}}$ multiplicity, as summarized
in \cref{tab:selection}. These channel definitions and selection
requirements follow \cite{ATLAS:2025eua}; entries marked ``---''
indicate that the corresponding requirement is not applied in that
channel.

\begin{table}[h]
\caption{Per-channel selection requirements following \cite{ATLAS:2025eua}.
SS = same sign}\label{tab:selection}
\begin{tabular*}{\textwidth}{@{\extracolsep\fill}lllllll}
\toprule
 & $2\ell$SS$0\tau$ & $3\ell0\tau$ & $4\ell$
 & $2\ell$SS$1\tau$ & $1\ell2\tau$ & $2\ell$$2\tau$ \\
\midrule
$N_\ell$          & 2      & 3      & $\geq$4 & 2      & 1      & 2      \\
$\sum q_\ell$     & $\pm2$ & $\pm1$ & 0       & $\pm2$ & $\pm1$ & 0      \\
$p_T(\ell_0)$ [GeV] & $>15$ & $>10$ & $>10$  & $>15$  & $>27$  & $>10$  \\
$N_\tau$          & 0      & 0      & ---     & 1      & 2      & 2      \\
$\tau\;p_T$ [GeV] & ---    & ---    & ---     & $>20$  & $>20$  & $>20$  \\
$N_\mathrm{jets}$ & $\geq3$& $\geq2$& $\geq2$ & $\geq4$& $\geq1$& $\geq1$\\
$N_b$             & $\geq1$& $\geq1$& $\geq1$ & $\geq1$& $\geq1$& $\geq1$\\
$|m_{\ell\ell}{-}m_Z|$ [GeV] & --- & $>10$ & --- & $>10$ & --- & $>10$ \\
$m_{\ell\ell}$ [GeV] & --- & $>12$ & $>12$  & ---    & ---    & $>12$  \\
\botrule
\end{tabular*}
\end{table}

\subsection{Feature hierarchy dimensions}\label{app:feature_summary}

\cref{tab:feature_dims} summarizes the input dimensionality at each
feature level for all six architectures. The same raw features are
made available to every architecture at a given level; how each
architecture actually consumes that information can differ by
design: e.g.\ MPNN+RoPE discards $p_x,p_y$ from its node encoder
inputs and instead computes $\phi$ internally for use in rotary
position encoding, ParT similarly strips $p_x,p_y$ (and, from
\fs{detector} onward, $\phi$) from its node inputs while retaining
azimuthal information via the pair bias, and LorentzNet is restricted
to Lorentz-invariant quantities and is therefore not evaluated at
\fs{detector} and \fs{full}.

\begin{table}[h]
\caption{Feature dimensions at each level for all architectures.
Node dim gives the raw per-particle feature count before any
architecture-specific stripping; Global dim gives the dimension of
the event-level input vector at \fs{full}, identical for flat and
graph models. Flat cols is the total input width for MLP and XGBoost
(Node dim $\times$ 14 slots, plus Global dim at \fs{full}). MPNN+RoPE
Node gives the effective node-encoder input width after discarding
$p_x,p_y$ (used instead for rotary position encoding, not as a node
feature). ParT Nodes and Pairs report the effective node and
pairwise bias dimensions after node stripping. LN scalar dim is the
number of Lorentz-invariant scalars passed to LorentzNet as node
attributes; \fs{detector} and \fs{full} are excluded from LorentzNet
entirely ($^\dagger$) as those features are not Lorentz-invariant.}
\label{tab:feature_dims}
\begin{tabular}{@{}lcccccccl@{}}
\toprule
Level & Node dim & Global dim & Flat cols & MPNN+RoPE Node & \multicolumn{2}{c}{ParT}
  & LN scalar \\
\cmidrule(lr){6-7}
 & & & & & Node & Pairs & \\
\midrule
\fs{core}     & 4 & 0 &  56 & 2 & 2 & 4 & 1 \\
\fs{object}   & 5 & 0 &  70 & 3 & 3 & 5 & 2 \\
\fs{detector} & 9 & 0 & 126 & 7 & 6 & 7 & 2$^\dagger$ \\
\fs{full}     & 9 & 6 & 132 & 7 & 6 & 7 & 2$^\dagger$ \\
\botrule
\multicolumn{8}{l}{$^\dagger$ The \fs{detector} and \fs{full} levels are not Lorentz-invariant and are not used.}
\end{tabular}
\end{table}

\clearpage

\section{Implementation details}\label{app:implementation}
\setcounter{table}{0}
\setcounter{figure}{0}
\setcounter{equation}{0}

\subsection{Architecture and symmetry properties}\label{app:arch}
\cref{tab:arch_props} summarises the symmetry and structural
properties of all six models. The rows are grouped into unconstrained
(MLP, XGBoost, MPNN) and symmetry-constrained (MPNN+RoPE, ParT,
LorentzNet) architectures, matching the two-panel figure layout used
throughout the results. Within the attention-based graph models,
HPO consistently selects TransformerConv for the vanilla MPNN, so
the MPNN$\,{\to}\,$MPNN+RoPE$\,{\to}\,$ParT progression forms a
near-controlled ablation: all three share the same core attention
mechanism, differing only in the successive addition of azimuthal
symmetry encoding and pairwise physics features.

\begin{table}[h]
\caption{Architectural and symmetry properties of all six models.
\checkmark~= property holds exactly by construction; $\times$~= not
enforced. ``Pair bias'' denotes ParT's physics-informed pairwise
attention input. Both ParT and MPNN+RoPE achieve
exact SO(2) invariance but via different mechanisms: pair bias plus
node stripping for ParT, RoPE plus node stripping for MPNN+RoPE.}
\label{tab:arch_props}
\begin{tabular}{@{}lccccc@{}}
\toprule
Model & Input & Perm.\ inv. & SO(2) & SO(3,1) & Pair bias \\
\midrule
MLP         & flat  & $\times$   & $\times$   & $\times$   & $\times$   \\
XGBoost     & flat  & $\times$   & $\times$   & $\times$   & $\times$   \\
MPNN        & graph & \checkmark & $\times$   & $\times$   & $\times$   \\
\midrule
MPNN+RoPE   & graph & \checkmark & \checkmark & $\times$   & $\times$   \\
ParT        & graph & \checkmark & \checkmark & $\times$   & \checkmark \\
LorentzNet  & graph & \checkmark & \checkmark & \checkmark & $\times$   \\
\botrule
\end{tabular}
\end{table}

\subsection{Hyperparameter search spaces}\label{app:hpo}

For each architecture, hyperparameters were tuned using Optuna's
TPE sampler~\cite{Akiba:2019optuna}, with study names encoding the
feature set under consideration. \cref{tab:search_spaces} lists the
full search space per model, including learning rate schedules,
architectural width/depth ranges, and regularisation terms. All
neural models share a common cosine learning-rate schedule with
linear warmup; the corresponding warmup and cosine-annealing
parameters are tuned jointly with the architecture-specific terms
below. Parameters marked as fixed were set from preliminary scans
and excluded from the HPO search to control its dimensionality.

\begin{table}[h]
\caption{Hyperparameter search spaces. All neural models additionally
tune weight decay ($[10^{-6},10^{-3}]$ log-uniform) and batch size
(\{512, 1024, 2048, 4096\} for flat models; \{1024, 2048, 4096\} for
ParT). $^\dagger$Constrained to head dimension $\geq32$.
$^\ddagger$Fixed implementation choice, not searched by HPO.
$^\S$Only searched when the relevant conv type or pooling is selected
(MPNN only). CLS pooling is excluded from MPNN+RoPE as the
single-frequency azimuthal encoding is not well-defined for the
class token's dense cross-attention.}

\label{tab:search_spaces}
\begin{tabular}{@{}llp{4.8cm}@{}}
\toprule
Model & Parameter & Range / choices \\
\midrule
\multirow{5}{*}{XGBoost}
  & \texttt{max\_depth}         & 3--30 \\
  & \texttt{learning\_rate}     & $[0.01,0.3]$ log-uniform \\
  & \texttt{subsample}          & $[0.6,1.0]$ \\
  & \texttt{reg\_alpha/lambda}  & $[10^{-8},10]$ log-uniform \\
  & \texttt{min\_child\_weight} & 1--10 \\
\midrule
\multirow{8}{*}{MLP}
  & \texttt{n\_layers}              & 1--6 \\
  & \texttt{hidden\_size}           & \{64,128,256,512,1024\} \\
  & \texttt{dropout}                & $[0.0,0.5]$ \\
  & \texttt{activation}             & \{gelu, relu, silu, elu\} \\
  & \texttt{lr}                     & $[10^{-5},3\times10^{-3}]$ log-uniform \\
  & \texttt{warmup\_epochs}         & \{5, 10, 20\} \\
  & \texttt{cosine\_t\_max}         & 50--200 \\
  & \texttt{cosine\_eta\_min\_frac} & $[10^{-3},10^{-1}]$ log-uniform \\
\midrule
\multirow{13}{*}{MPNN}
  & \texttt{conv\_type}             & \{gcn, gat, sage, transformer\} \\
  & \texttt{hidden\_dim}            & \{64,128,256,512\} \\
  & \texttt{num\_layers}            & 2--6 \\
  & \texttt{pooling}                & \{mean, max, add, attention, cls\} \\
  & \texttt{activation}             & \{gelu, relu, silu, elu\} \\
  & \texttt{lr}                     & $[5\times10^{-5},3\times10^{-3}]$ log-uniform \\
  & \texttt{num\_heads}$^\S$        & \{4,8,16\} (gat, transformer) \\
  & \texttt{ffn\_multiplier}$^\S$   & \{2,4\} (transformer only) \\
  & \texttt{num\_cls\_layers}$^\S$  & 1--3 (cls pooling only) \\
  & \texttt{cls\_num\_heads}$^\S$   & \{4,8\} (cls pooling only) \\
  & \texttt{warmup\_epochs}         & \{5, 10, 20\} \\
  & \texttt{cosine\_t\_max}         & 50--200 \\
  & \texttt{cosine\_eta\_min\_frac} & $[10^{-3},10^{-1}]$ log-uniform \\
\midrule
\multirow{14}{*}{MPNN+RoPE}
  & \texttt{conv\_type}             & fixed: transformer$^\ddagger$ \\
  & \texttt{hidden\_dim}            & \{64,128,256,512\} \\
  & \texttt{num\_layers}            & 2--6 \\
  & \texttt{pooling}                & \{mean, max, add, attention\} \\
  & \texttt{num\_heads}             & \{4,8\} \\
  & \texttt{ffn\_multiplier}        & \{2,4\} \\
  & \texttt{attention\_dropout}     & $[0.0,0.4]$ \\
  & \texttt{conv\_dropout}          & $[0.0,0.5]$ \\
  & \texttt{final\_dropout}         & $[0.0,0.5]$ \\
  & \texttt{activation}             & \{gelu, relu, silu, elu\} \\
  & \texttt{lr}                     & $[5\times10^{-5},3\times10^{-3}]$ log-uniform \\
  & \texttt{warmup\_epochs}         & \{5, 10, 20\} \\
  & \texttt{cosine\_t\_max}         & 50--200 \\
  & \texttt{cosine\_eta\_min\_frac} & $[10^{-3},10^{-1}]$ log-uniform \\
  & \texttt{rope\_omega}            & fixed: $\omega=1$$^\ddagger$ \\
\midrule
\multirow{12}{*}{ParT}
  & \texttt{hidden\_dim}            & \{128,256,512\} \\
  & \texttt{num\_heads}$^\dagger$   & \{4,8\} \\
  & \texttt{num\_layers}            & 3--8 \\
  & \texttt{num\_cls\_layers}       & 1--3 \\
  & \texttt{ffn\_multiplier}        & \{2,4\} \\
  & \texttt{pair\_embed\_dims}      & fixed: $[64,64,64]$$^\ddagger$ \\
  & \texttt{attn\_dropout}          & $[0.0,0.3]$ \\
  & \texttt{ffn\_dropout}           & $[0.0,0.3]$ \\
  & \texttt{lr}                     & $[10^{-5},2\times10^{-3}]$ log-uniform \\
  & \texttt{warmup\_epochs}         & \{5, 10, 20\} \\
  & \texttt{cosine\_t\_max}         & 50--200 \\
  & \texttt{cosine\_eta\_min\_frac} & $[10^{-3},10^{-1}]$ log-uniform \\
\midrule
\multirow{9}{*}{LorentzNet}
  & \texttt{hidden\_dim}            & \{64,72,128,192,256\} \\
  & \texttt{num\_layers}            & 2--8 \\
  & \texttt{c\_weight}              & $[10^{-4},10^{-1}]$ log-uniform \\
  & \texttt{dropout}                & $[0.0,0.5]$ \\
  & \texttt{pooling}                & \{mean, max, add\} \\
  & \texttt{lr}                     & $[10^{-5},3\times10^{-3}]$ log-uniform \\
  & \texttt{warmup\_epochs}         & 0--40 \\
  & \texttt{cosine\_t\_max}         & 50--200 \\
  & \texttt{cosine\_eta\_min\_frac} & $[10^{-3},10^{-1}]$ log-uniform \\
\botrule
\end{tabular}
\end{table}
\clearpage

\subsection{Training procedure}\label{app:train}

Four NVIDIA RTX 2080 Ti GPUs (11~GB VRAM each) run trials in parallel.
The number of trials varies by architecture, reflecting differences in
search space size and per-trial cost; optimisation is stopped when no
improvement is observed over at least 15 consecutive trials (usually around 40-50 in total). To maximise throughput, each neural HPO trial uses a
reduced epoch budget of 150 epochs and patience 15; relative ordering
of configurations is stable well before convergence, as suggested by
inspection of intermediate validation curves.

All five neural models (MLP, MPNN, MPNN+RoPE, ParT, LorentzNet) use
AdamW optimisation with linear warmup followed by cosine annealing;
the warmup length, cosine period $T_\mathrm{max}$, and minimum
learning rate fraction $\eta_\mathrm{min}/\eta_0$ are tuned by HPO
for every architecture (\cref{app:hpo}). Experiments use
Python~3.11, PyTorch~2.2, PyTorch Geometric~2.5, XGBoost~2.0, and
Optuna~3.5.

\section{Extended results and Verification}\label{app:results}
\setcounter{table}{0}
\setcounter{figure}{0}
\setcounter{equation}{0}

\subsection{Feature test}\label{app:feature}

\cref{tab:features} gives the exact values underlying
\cref{fig:feature_auc} (ROC AUC) and the corresponding $\varrho_\mu$
results discussed in \cref{sec:feature}.

\begin{table*}[htbp]
\centering
\caption{Feature experiment: dependence on the feature-set level,
averaged over five seeds. $\varrho_\mu$ is evaluated at
$3000~\mathrm{fb}^{-1}$. LorentzNet is evaluated only at the
\fs{core} and \fs{object} levels, as it is not applicable to the
\fs{detector} and \fs{full} feature sets by design.}
\label{tab:features}
\textbf{(a) ROC AUC}\par\vspace{1mm}
\begin{tabular}{lcccc}
\toprule
Model & core & object & detector & full \\
\midrule
MLP         & 0.7497 & 0.7527 & 0.7575 & 0.7627 \\
XGBoost     & 0.7582 & 0.7628 & 0.7694 & 0.7712 \\
MPNN        & 0.7677 & 0.7717 & 0.7769 & 0.7784 \\
MPNN+RoPE   & 0.7711 & 0.7743 & 0.7829 & 0.7857 \\
ParT        & 0.7663 & 0.7788 & 0.7865 & 0.7982 \\
LorentzNet  & 0.7890 & 0.7991 & --     & --     \\
\bottomrule
\end{tabular}\par
\vspace{3mm}
\textbf{(b) $\varrho_\mu$ at 3000 $fb^{-1}$}\par\vspace{1mm}
\begin{tabular}{lcccc}
\toprule
Model & core & object & detector & full \\
\midrule
MLP         & 0.8697 & 0.8694 & 0.8591 & 0.8535 \\
XGBoost     & 0.8538 & 0.8502 & 0.8434 & 0.8412 \\
MPNN        & 0.8375 & 0.8340 & 0.8231 & 0.8236 \\
MPNN+RoPE   & 0.8302 & 0.8285 & 0.8123 & 0.8121 \\
ParT        & 0.8315 & 0.8135 & 0.7996 & 0.7911 \\
LorentzNet  & 0.7982 & 0.7844 & --     & --     \\
\bottomrule
\end{tabular}\par
\end{table*}

\clearpage

\subsection{Scaling test}\label{app:scale}
\cref{tab:scaling} gives the exact values underlying
\cref{fig:scaling_auc} (ROC AUC) and the corresponding $\varrho_\mu$
results discussed in \cref{sec:scaling}.
\begin{table*}[htbp]
\centering
\caption{Scaling experiment: dependence on the fraction of training
data used, averaged over five seeds. $\varrho_\mu$ is evaluated at
$3000~\mathrm{fb}^{-1}$.}
\label{tab:scaling}
\begin{center}
\textbf{(a) ROC AUC}\\[1mm]
\begin{tabular}{lcccccc}
\toprule
Model & 10\% & 20\% & 40\% & 60\% & 80\% & 100\% \\
\midrule
MLP         & 0.7411 & 0.7486 & 0.7527 & 0.7576 & 0.7611 & 0.7632 \\
XGBoost     & 0.7564 & 0.7620 & 0.7653 & 0.7677 & 0.7698 & 0.7710 \\
MPNN        & 0.7542 & 0.7594 & 0.7670 & 0.7705 & 0.7753 & 0.7794 \\
MPNN+RoPE   & 0.7612 & 0.7683 & 0.7728 & 0.7780 & 0.7821 & 0.7859 \\
ParT        & 0.7601 & 0.7685 & 0.7834 & 0.7903 & 0.7947 & 0.7986 \\
LorentzNet  & 0.7786 & 0.7865 & 0.7914 & 0.7950 & 0.7969 & 0.7987 \\
\bottomrule
\end{tabular}
\end{center}
\vspace{3mm}
\begin{center}
\textbf{(b) $\varrho_\mu$ at 3000 $fb^{-1}$}\\[1mm]
\begin{tabular}{lcccccc}
\toprule
Model & 10\% & 20\% & 40\% & 60\% & 80\% & 100\% \\
\midrule
MLP         & 0.8841 & 0.8740 & 0.8673 & 0.8593 & 0.8552 & 0.8515 \\
XGBoost     & 0.8621 & 0.8548 & 0.8495 & 0.8462 & 0.8437 & 0.8410 \\
MPNN        & 0.8603 & 0.8516 & 0.8422 & 0.8363 & 0.8308 & 0.8225 \\
MPNN+RoPE   & 0.8476 & 0.8380 & 0.8320 & 0.8238 & 0.8184 & 0.8121 \\
ParT        & 0.8540 & 0.8411 & 0.8176 & 0.8068 & 0.7970 & 0.7895 \\
LorentzNet  & 0.8214 & 0.8061 & 0.7965 & 0.7910 & 0.7882 & 0.7845 \\
\bottomrule
\end{tabular}
\end{center}

\end{table*}

\subsection{Joint versus dedicated training}\label{app:joint}

\cref{tab:channel_res_final} gives the exact values underlying
\cref{fig:channel_auc,fig:channel_statunc} (ROC AUC) and the corresponding $\varrho_\mu $
results discussed in \cref{sec:channels}.

\begin{table}[ht]
\caption{Per-channel ROC AUC and statistical uncertainty ratio
$\varrho_\mu$ (relative to the 1-bin reference, at
$\mathcal{L}=3000~\mathrm{fb}^{-1}$; lower is better) at the highest feature level. ``Combined''
denotes multi-channel training; ``Dedicated'' denotes single-channel
training. Values are means over five seeds. For each model, the
better value between training modes is \textbf{bolded} (higher for
AUC, lower for $\varrho_\mu$).}
\label{tab:channel_res_final}
\small
\centering
\begin{tabular}{@{}llcccccc@{}}
\toprule
Model & Mode
  & $2\ell\mathrm{SS}0\tau$
  & $3\ell0\tau$
  & $4\ell$
  & $2\ell\mathrm{SS}1\tau$
  & $1\ell2\tau$
  & $2\ell2\tau$ \\
\midrule
\multicolumn{8}{c}{\textit{ROC AUC}} \\
\midrule
MLP
  & Combined  & \textbf{0.735} & \textbf{0.776} & 0.704
              & \textbf{0.681} & \textbf{0.608} & \textbf{0.619} \\
  & Dedicated & 0.732 & 0.762 & \textbf{0.704} & 0.660 & 0.595 & 0.528 \\
\addlinespace
XGBoost
  & Combined  & 0.743 & \textbf{0.787} & 0.718
              & \textbf{0.720} & \textbf{0.629} & \textbf{0.598} \\
  & Dedicated & \textbf{0.743} & 0.782 & \textbf{0.736} & 0.694 & 0.619 & 0.572 \\
\addlinespace
MPNN
  & Combined  & \textbf{0.750} & \textbf{0.796} & \textbf{0.790}
              & \textbf{0.717} & \textbf{0.648} & \textbf{0.629} \\
  & Dedicated & 0.744 & 0.780 & 0.727 & 0.677 & 0.600 & 0.555 \\
\addlinespace
MPNN+RoPE
  & Combined  & \textbf{0.758} & \textbf{0.802} & \textbf{0.806}
              & \textbf{0.717} & \textbf{0.668} & \textbf{0.639} \\
  & Dedicated & 0.750 & 0.787 & 0.747 & 0.687 & 0.604 & 0.587 \\
\addlinespace
ParT
  & Combined  & \textbf{0.767} & \textbf{0.818} & \textbf{0.914}
              & \textbf{0.734} & \textbf{0.703} & \textbf{0.701} \\
  & Dedicated & 0.758 & 0.798 & 0.771 & 0.687 & 0.647 & 0.556 \\
\addlinespace
LorentzNet
  & Combined  & \textbf{0.763} & \textbf{0.815} & \textbf{0.941}
              & \textbf{0.747} & \textbf{0.711} & \textbf{0.721} \\
  & Dedicated & 0.762 & 0.804 & 0.884 & 0.718 & 0.701 & 0.593 \\
\midrule
\multicolumn{8}{c}{\textit{$\varrho_\mu$ at $3000~\mathrm{fb}^{-1}$}} \\
\midrule
MLP
  & Combined  & \textbf{0.804} & \textbf{0.807} & 0.770
              & \textbf{0.910} & \textbf{0.966} & \textbf{0.921} \\
  & Dedicated & 0.811 & 0.825 & \textbf{0.762} & 0.922 & 0.975 & 0.956 \\
\addlinespace
XGBoost
  & Combined  & \textbf{0.796} & \textbf{0.796} & 0.729
              & \textbf{0.881} & \textbf{0.961} & \textbf{0.919} \\
  & Dedicated & 0.796 & 0.803 & \textbf{0.677} & 0.907 & 0.970 & 0.950 \\
\addlinespace
MPNN
  & Combined  & \textbf{0.774} & \textbf{0.781} & \textbf{0.600}
              & \textbf{0.876} & \textbf{0.943} & \textbf{0.912} \\
  & Dedicated & 0.782 & 0.799 & 0.731 & 0.913 & 0.972 & 0.955 \\
\addlinespace
MPNN+RoPE
  & Combined  & \textbf{0.762} & \textbf{0.774} & \textbf{0.581}
              & \textbf{0.881} & \textbf{0.924} & \textbf{0.901} \\
  & Dedicated & 0.773 & 0.789 & 0.676 & 0.903 & 0.971 & 0.932 \\
\addlinespace
ParT
  & Combined  & \textbf{0.751} & \textbf{0.752} & \textbf{0.390}
              & \textbf{0.862} & \textbf{0.891} & \textbf{0.892} \\
  & Dedicated & 0.769 & 0.778 & 0.655 & 0.906 & 0.940 & 0.936 \\
\addlinespace
LorentzNet
  & Combined  & \textbf{0.751} & \textbf{0.755} & \textbf{0.351}
              & \textbf{0.846} & \textbf{0.879} & \textbf{0.883} \\
  & Dedicated & 0.754 & 0.769 & 0.459 & 0.875 & 0.886 & 0.932 \\
\botrule

\multicolumn{8}{l}{}
\end{tabular}
\end{table}

\clearpage

\subsection{Azimuthal rotation test}\label{app:phi}

The test applies 12 equally spaced global $\phi$ rotations,
$\Delta\phi_k = k\pi/6$ for $k = 0,\dots,11$, to every event in the
test set by replacing $\phi_i \to \phi_i + \Delta\phi_k \pmod{2\pi}$
and updating the Cartesian momentum components consistently. The test
is applied at the highest feature level using models trained on
100\% of the data; no retraining is required.

\begin{figure*}[!ht]
\centering
\includegraphics[width=0.49\textwidth]{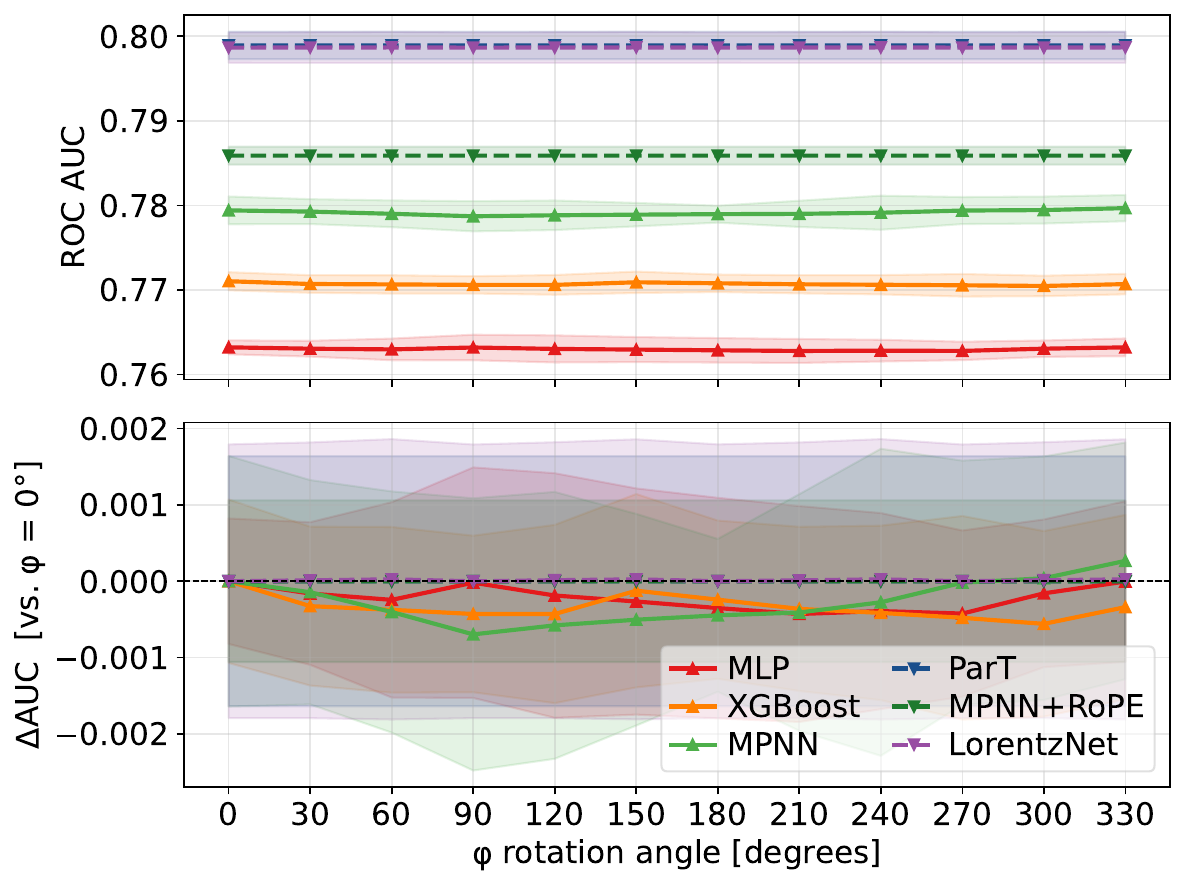}
\includegraphics[width=0.49\textwidth]{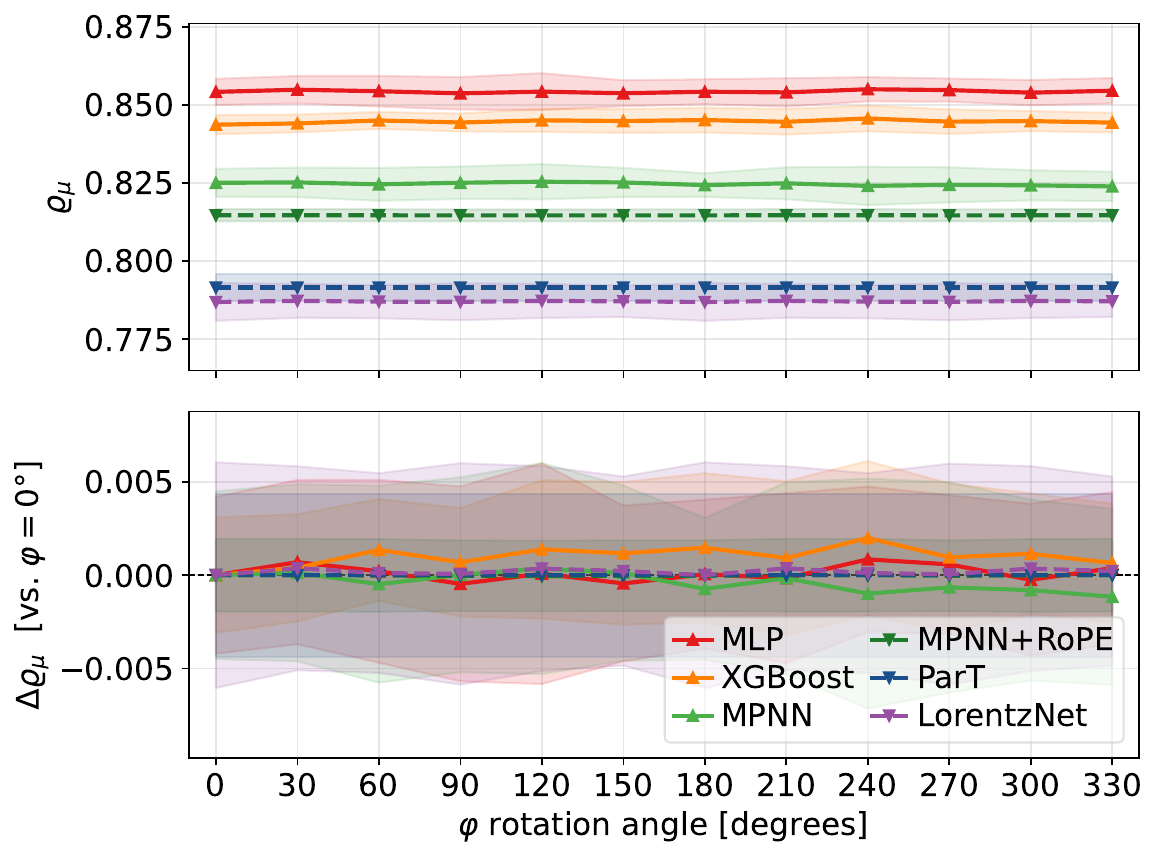}
\caption{Sensitivity to global azimuthal rotation angle
$\Delta\phi_k = k\pi/6$ ($k=0,\dots,11$) applied to all particles
simultaneously on the held-out test set at the highest feature
level, 100\% training data. \emph{(Left)}: ROC AUC; left panel shows
unconstrained models (MLP, XGBoost, MPNN), right panel shows
symmetry-constrained models (MPNN+RoPE, ParT, LorentzNet), with grey
dashed lines reproducing MPNN for reference. \emph{(Right)}: profile
likelihood ratio $\sigma(\mu)/\sigma_\mathrm{ref}$ at the HL-LHC
projection luminosity.}
\label{fig:phi_rotation}
\end{figure*}

Both MPNN+RoPE and ParT achieve exactly zero AUC variation across all
rotation angles (\cref{fig:phi_rotation}), confirming exact
azimuthal invariance. LorentzNet shows a small residual, attributable
to \texttt{float32} accumulation of rounding errors in the Minkowski
inner products rather than a theoretical breakdown of invariance.
Unconstrained models show consistent non-zero sensitivity, with MPNN
most affected, followed by XGBoost and MLP.

In the profile likelihood metric at $3000~\mathrm{fb}^{-1}$
(\cref{fig:phi_rotation}), residuals are larger than in AUC,
reflecting the amplification of score distribution shape differences
by the likelihood fit. ParT remains exactly flat. MPNN+RoPE shows a
negligible residual, and LorentzNet a somewhat larger one, consistent
with floating-point evaluation of the \texttt{TransfoD} binning
boundaries shifting marginally under rotation. All constrained
residuals remain substantially smaller than those of
unconstrained architectures and are negligible in practice. Full
numerical values are summarised in \cref{tab:phi}.

\begin{table}[h]
\caption{Azimuthal rotation sensitivity at the \fs{full} feature
level, 100\% training data, $\mathcal{L}=3000~\mathrm{fb}^{-1}$.
$\sigma(\mathrm{AUC})$ and $\sigma(\varrho_\mu)$ are the standard
deviations of ROC AUC and $\varrho_\mu$ respectively, over 12 equally
spaced global $\phi$ rotations; max $|$deviation$|$ is the largest
absolute difference from the $\phi=0^\circ$ value in each metric.
Lower is better in all columns.}
\label{tab:phi}
\begin{tabular}{@{}lcccc@{}}
\toprule
& \multicolumn{2}{c}{ROC AUC} & \multicolumn{2}{c}{$\varrho_\mu$} \\
\cmidrule(lr){2-3}\cmidrule(lr){4-5}
Model & $\sigma(\mathrm{AUC})$ & max $|$dev.$|$ &
  $\sigma(\varrho_\mu)$ & max $|$dev.$|$ \\
\midrule
\multicolumn{5}{l}{\textit{Unconstrained}} \\
MLP        & $0.000152$ & $0.000430$ & $0.000429$ & $0.000844$ \\
XGBoost    & $0.000148$ & $0.000559$ & $0.000530$ & $0.001980$ \\
MPNN       & $0.000279$ & $0.000697$ & $0.000498$ & $0.001153$ \\
\midrule
\multicolumn{5}{l}{\textit{Symmetry-constrained}} \\
MPNN+RoPE  & $0.000000$ & $0.000000$ & $0.000016$ & $0.000047$ \\
ParT       & $0.000000$ & $0.000000$ & $0.000000$ & $0.000000$ \\
LorentzNet & $0.000010$ & $0.000025$ & $0.000146$ & $0.000366$ \\
\botrule
\end{tabular}
\end{table}

\subsection{Luminosity scan robustness}
\label{app:lumi-scan}

The main text reports $\varrho_\mu$ at a single HL-LHC luminosity
point, $\mathcal{L}=3000$~fb$^{-1}$. To verify that the architecture
ranking observed there generalises rather than being specific to that
one luminosity, a scan over $\mathcal{L}\in[10,5000]$~fb$^{-1}$ is
performed for all six architectures at the highest feature level, with the \texttt{TransfoD} parameter
fixed at its default value ($a=10$, Appendix~\ref{app:transfod-scan})
and the full TRExFitter fit re-run at each luminosity point.

\begin{figure}[t]
\centering
\includegraphics[width=\linewidth]{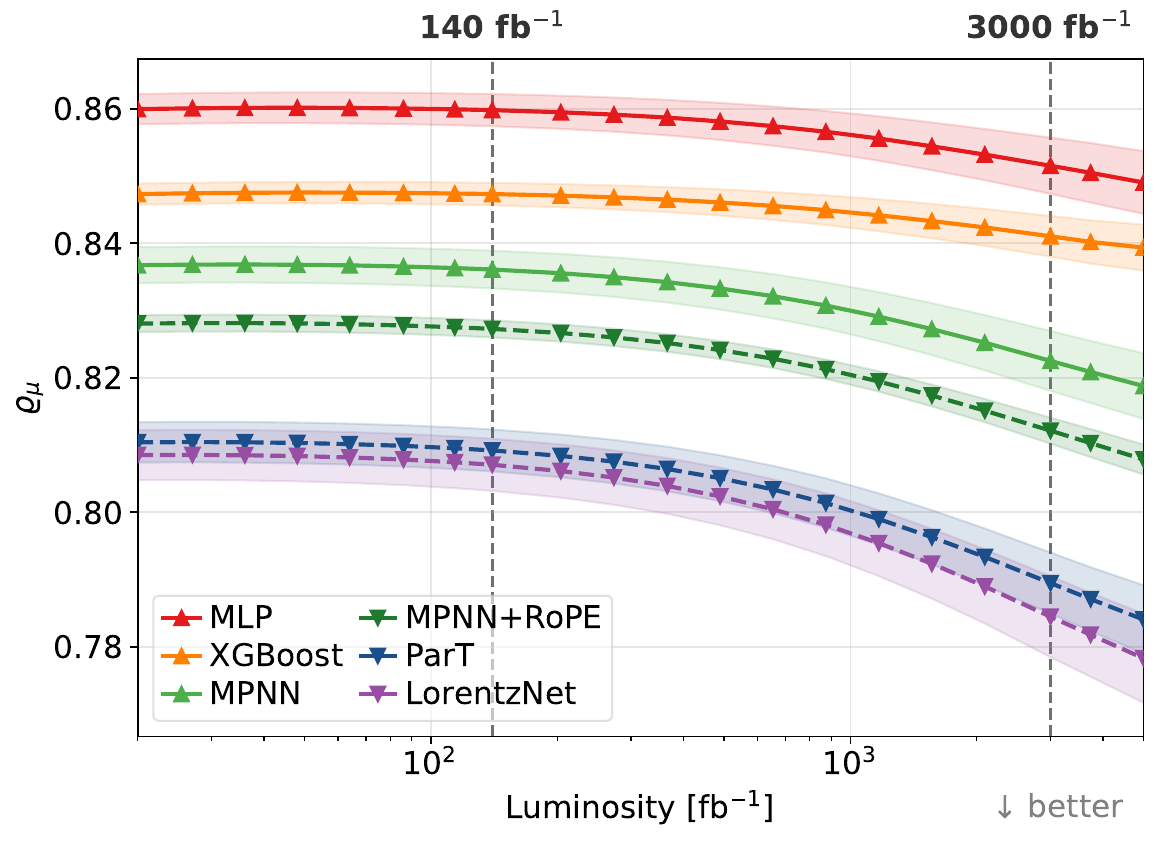}
\caption{$\varrho_\mu$ vs.\ luminosity from $10$ to $5000$~fb$^{-1}$ (20 points),
all six architectures at the highest feature level,
\texttt{TransfoD} parameter fixed at $a=10$. Dashed lines mark
$140$~fb$^{-1}$, the integrated luminosity of the finalized ATLAS/CMS
Run-2 (2015--2018) dataset, and $3000$~fb$^{-1}$, the nominal HL-LHC
integrated luminosity used throughout the main results. The relative
ranking of architectures is unchanged across the entire scanned range.
Shaded bands: standard deviation over five seeds.}
\label{fig:lumi_scan}
\end{figure}

Figure~\ref{fig:lumi_scan} shows that the relative ordering of all six
architectures is unchanged across the full scanned range, confirming
that the ranking reported at $3000$~fb$^{-1}$ is not a special or
isolated result but holds generally across luminosities. The origin of the shape has been determined to arise from the combination of statistical and systematic uncertainties, where systematic uncertainties are due to the limited number of simulated events.

\subsection{\texttt{TransfoD} binning robustness scan}
\label{app:transfod-scan}

To verify that the reported $\varrho_\mu$ values are not an artefact
of the \texttt{TransfoD} binning parameter used throughout the paper
($a\equiv z_b=z_s=10$, \cref{sec:statistics}), a scan over
$a\in[1,20]$ is performed at $\mathscr{L}=3000$~fb$^{-1}$ for all six
architectures at the highest feature level, re-running the full TRExFitter fit at each value.

\begin{figure}[t]
\centering
\includegraphics[width=\linewidth]{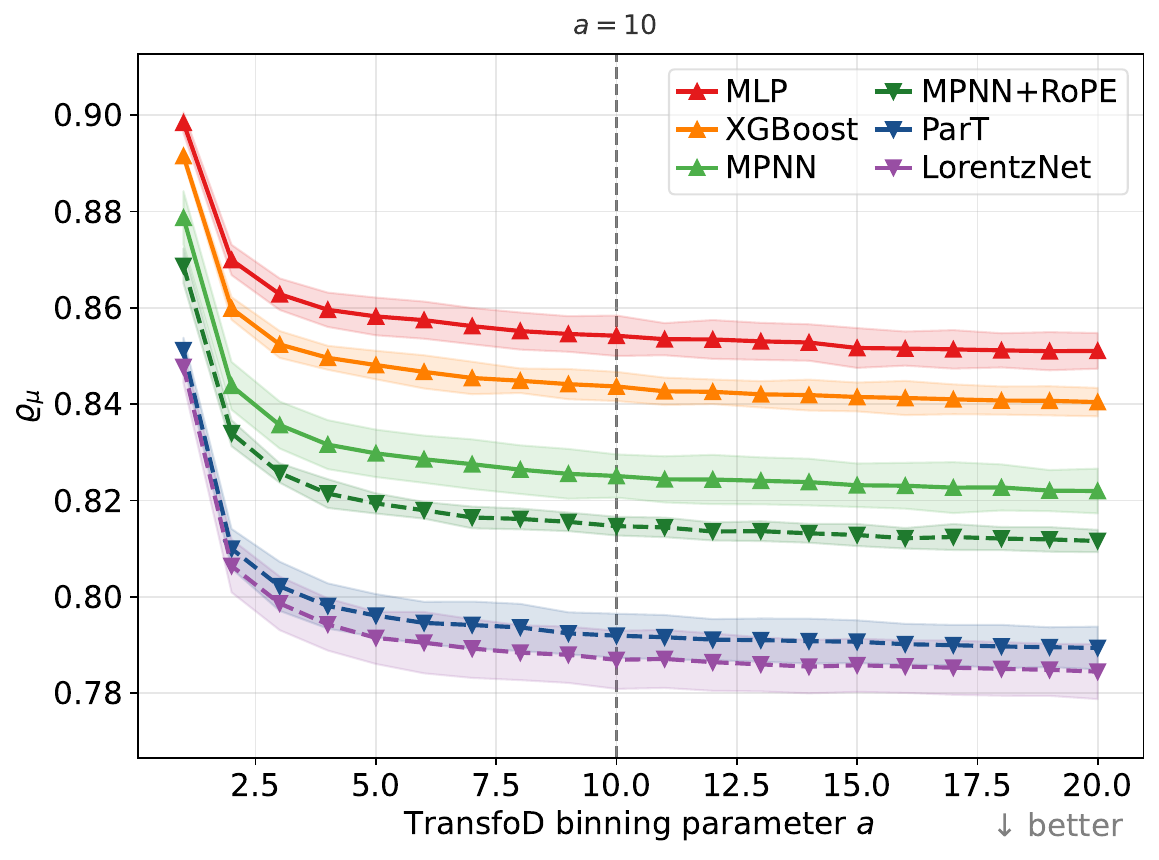}
\caption{$\varrho_\mu$ vs.\ \texttt{TransfoD} parameter $a$
($z_b=z_s=a$) at $\mathscr{L}=3000$~fb$^{-1}$ from 1 to 20 (20 points), all six architectures
at the highest feature level. Values plateau by
$a\gtrsim10$; the default $a=10$ (dashed line) lies within this
plateau, and architecture ranking is unchanged across the scanned
range. Shaded bands: standard deviation over five seeds.}
\label{fig:transfod_scan}
\end{figure}

Smaller $a$ means finer binning and fewer effective Monte-Carlo events
per bin; $\varrho_\mu$ falls steeply
with increasing $a$ before flattening into a stable plateau by
$a\gtrsim10$. \cref{fig:transfod_scan} shows the default $a=10$
lies within this plateau for every architecture, and the relative
ranking is unchanged across the full scanned range.

\clearpage

\subsection{Summary table}\label{app:summary}

\cref{tab:summary} collects the headline numbers referenced in the
Conclusion (\cref{sec:conclusion}): ROC AUC and $\varrho_\mu$ for all
six architectures at the highest feature level and 100\% of the
training data, the configuration under which each model attains its
best overall performance in this benchmark.

\begin{table}[ht]
\caption{Summary of model performance at the highest feature level,
100\% training data, $\mathcal{L}=3000~\mathrm{fb}^{-1}$. Values are
mean $\pm$ standard deviation over five seeds. Best value in each
column is \textbf{bolded}.}
\label{tab:summary}
\small
\centering
\begin{tabular}{@{}lcc@{}}
\toprule
Model
  & ROC AUC
  & $\sigma(\mu)/\sigma_\mathrm{ref}$ \\
\midrule
\multicolumn{3}{l}{\textit{Unconstrained}} \\
MLP        & $0.7632 \pm 0.0009$ & $0.852 \pm 0.004$ \\
XGBoost    & $0.7710 \pm 0.0012$ & $0.841 \pm 0.003$ \\
MPNN       & $0.7794 \pm 0.0018$ & $0.823 \pm 0.004$ \\
\midrule
\multicolumn{3}{l}{\textit{Symmetry-constrained}} \\
MPNN+RoPE  & $0.7859 \pm 0.0012$ & $0.812 \pm 0.002$ \\
ParT       & $0.7986 \pm 0.0018$ & $0.789 \pm 0.005$ \\
LorentzNet
           & $\mathbf{0.7987 \pm 0.0020}$ & $\mathbf{0.784 \pm 0.006}$ \\
\botrule
\multicolumn{3}{l}{}
\end{tabular}
\end{table}

\section{Statistical model details}\label{app:stat-model}

This appendix briefly summarises the explicit form of the
per-bin expected yield, the Barlow–Beeston--lite Monte-Carlo
statistical nuisance, and the resulting luminosity scaling of
$\sigma(\mu)$, in order to see how the process weights enter the likelihood, how the scaling behaviour discussed above arises, and how the analytic solution for $\sigma(\mu)$ is obtained.

\subsection{Expected bin yields}\label{app:stat_rate}

Refining the total preselection efficiency
$\varepsilon_p=N_{\mathrm{sel},p}/N_{\mathrm{gen},p}$ of the per-bin level,
$\varepsilon_{ck}^{\,p}=N^{\mathrm{MC}}_{ck,p}/N_{\mathrm{gen},p}$ is
the fraction of generated events of process $p$ landing in bin
$(c,k)$, with $\sum_{c,k}\varepsilon_{ck}^{\,p}=\varepsilon_p$. At luminosity  $\mathscr{L}$ the
expected yield in bin $(c,k)$ is then
\begin{equation}
\nu_{ck}(\mu,\boldsymbol{\theta}) = \gamma_{ck}\cdot \mathscr{L} \left[
\mu \cdot \sigma_{t\bar{t}H}\,\varepsilon_{ck}^{\,t\bar{t}H}
+ \sigma_{t\bar{t}W}\,\varepsilon_{ck}^{\,t\bar{t}W}
+ \sigma_{t\bar{t}Z}\,\varepsilon_{ck}^{\,t\bar{t}Z}
\right],
\label{eq:nu_ck}
\end{equation}
with $\boldsymbol{\theta} = \{\gamma_{ck}\}$: $\mu$ scales only the \tth\ term, consistent with
its role as the parameter of interest; the $t\bar{t}W$ and
$t\bar{t}Z$ terms are fixed to their Standard Model predictions
throughout this study (main fit, six-channel-bin reference, and
individual-channel reference alike) rather than floated; and $\gamma_{ck}$ is the
pooled Barlow--Beeston-lite Monte-Carlo statistical nuisance multiplying the \emph{total} bin rate, discussed next.

\subsection{Monte-Carlo statistical nuisance}\label{app:stat_gamma}

The constraint $\mathrm{Gam}(\gamma_{ck})$ in \cref{eq:likelihood} is equivalent to \texttt{autoMCStats}
implementation~\cite{Cranmer:2012sba} of the per-bin Monte-Carlo
statistical uncertainty: Barlow--Beeston lite~\cite{Conway:2011in}
pools all processes in a bin into a single nuisance, rather than one
nuisance per process per bin as in the original Barlow--Beeston
method~\cite{Barlow:1993dm}. Writing
$\hat\nu_{ck}$ for the pooled nominal prediction in bin $(c,k)$
(Eq.~\eqref{eq:nu_ck} at $\gamma_{ck}=1$)
and $\sigma^2_{\mathrm{MC},ck}$ for the corresponding sum of squared
per-event Monte-Carlo weights in that bin, the probabilistic constraint is
\begin{equation}
\mathrm{Gam}(\gamma_{ck};\,\tau_{ck}+1,\,\tau_{ck}), \qquad
\tau_{ck} = \hat\nu_{ck}^2 / \sigma^2_{\mathrm{MC},ck},
\label{eq:tau_def}
\end{equation}
with mode $1$ (the nominal value) and, for the large $\tau_{ck}$
typical here, variance $\sigma^2_{\mathrm{MC},ck}/\hat\nu_{ck}^2$.
$\tau_{ck}$ is the \emph{effective number of Monte Carlo events}
pooled into bin $(c,k)$, a property of the simulated sample w.r.t. the binning: it
does not depend on $\mathscr{L}$, since both $\hat\nu_{ck}$ and
$\sigma_{\mathrm{MC},ck}$ scale by the same overall factor with $\mathscr{L}$.

\subsection{Luminosity scaling of \texorpdfstring{$\sigma(\mu)$}{sigma(mu)}}\label{app:stat_scaling}
The signal uncertainty $\sigma(\mu)$ is the inverse curvature of the profile likelihood at
the Asimov point $\mu=1$ in a Gaussian approximation~\cite{Cowan:2010js}: the standard deviation
of $\hat\mu$ over hypothetical repeated measurements at luminosity
$\mathscr{L}$, exact in the asymptotic (Wald) regime, where it coincides with
$\mu/Z$ for the Asimov significance $Z$ from the profile-likelihood
test statistic $q_\mu$~\cite{Cowan:2010js}. Because $k_{t\bar{t}W}=k_{t\bar{t}Z}=1$
are fixed throughout this study, the reference uncertainty for a
single channel, $\sigma_{\mathrm{ref}}^{c}$, admits a closed-form
solution below; the multi-channel reference $\sigma_{\mathrm{ref}}$
(\cref{sec:statistics}) is then simply the inverse-variance
combination of the six per-channel results.

For each channel $c$, $\sigma_{\mathrm{ref}}^{c}$ is the uncertainty from a
single bin containing only events from that channel. With the
per-process weight
\begin{equation}
w_p = \frac{\sigma_p}{N_{\mathrm{gen},p}\cdot f_{\mathrm{test}}}, \qquad f_{\mathrm{test}}=0.10,
\label{eq:weights_test}
\end{equation}
where $f_{\mathrm{test}}$ is the fraction of generated events retained
in the held-out test split used for
evaluation, so that $N_{\mathrm{test},p}^{c}$ below is counted directly
from that test-split sample rather than extrapolated to the full
selected sample, the nominal signal and total yields of
channel $c$ at luminosity $\mathscr{L}$ are
\begin{equation}
s_c = \mathscr{L}\,w_{t\bar{t}H}\,N_{\mathrm{test},t\bar{t}H}^{c},\qquad
\nu_{0,c} = \mathscr{L}\sum_p w_p\,N_{\mathrm{test},p}^{c},
\end{equation}
the sum running over the three simulated processes and
$N_{\mathrm{test},p}^{c}$ the number of selected events of process
$p$ in channel $c$ within the test-split sample. The summed squared Monte-Carlo weights are
$\sigma^2_{\mathrm{MC},c}=\mathscr{L}^2\sum_p w_p^2\,N_{\mathrm{test},p}^{c}$,
so the effective count $\tau_c=\nu_{0,c}^2/\sigma^2_{\mathrm{MC},c}$ is
$\mathscr{L}$-independent (\cref{app:stat_gamma}). The Asimov Fisher
information in $(\mu,\gamma_c)$ at $\mu=\gamma_c=1$ is
\begin{equation}
I_c=\begin{pmatrix} s_c^2/\nu_{0,c} & s_c\\ s_c & \nu_{0,c}+\tau_c\end{pmatrix},
\end{equation}
where the Poisson part being rank one with the $\gamma_c$ constraint adding
$\tau_c$ on its diagonal. Profiling out $\gamma_c$ by the Schur
complement, $\tilde I_{\mu\mu}=s_c^2/(\nu_{0,c}+\sigma^2_{\mathrm{MC},c})$,
gives
\begin{equation}
\left(\sigma_{\mathrm{ref}}^{c}\right)^2=\frac{\nu_{0,c}+\sigma^2_{\mathrm{MC},c}}{s_c^2}
=\underbrace{\frac{\sum_p w_p N_{\mathrm{test},p}^{c}}{\mathscr{L}\,(w_{t\bar{t}H}N_{\mathrm{test},t\bar{t}H}^{c})^2}}_{\text{data statistics}}
+\underbrace{\frac{\sum_p w_p^2 N_{\mathrm{test},p}^{c}}{(w_{t\bar{t}H}N_{\mathrm{test},t\bar{t}H}^{c})^2}}_{\text{Monte-Carlo statistics}}.
\label{eq:sigma_scaling_channel}
\end{equation}
The data-Poisson term falls as $1/\mathscr{L}$ while the
Monte-Carlo term is an $\mathscr{L}$-independent floor set by simulated sample size of  channel $c$, the two crossing over at
$\mathscr{L}_c^\star=\sum_p w_p N_{\mathrm{test},p}^{c}\big/\sum_p w_p^2 N_{\mathrm{test},p}^{c}$.

For the multi-channel studies,
$\sigma_{\mathrm{ref}}$ is defined from the same six channels used in
the simultaneous \TREx\ fit for $\sigma(\mu)$: same samples, same
fixed $k_{t\bar{t}W}$, $k_{t\bar{t}Z}$ at their Standard Model
values, same Barlow--Beeston-lite treatment of $\gamma_{ck}$, but with
each channel collapsed to a single bin rather than binned by
classifier score with \texttt{TransfoD}. Because the six channels are
statistically independent and share only $\mu$, with no correlated
systematic linking them, their Asimov Fisher information simply adds,
and $\sigma_{\mathrm{ref}}$ follows by combining the six closed-form
channel results of Eq.~\eqref{eq:sigma_scaling_channel} via
inverse-variance weighting, Eq.~\eqref{eq:sigma_ref_combined} of
\cref{sec:statistics}.

The data-Poisson variance of each bin scales as $\mathscr{L}$
and its Monte-Carlo variance as $\mathscr{L}^2$ (\cref{app:stat_gamma}),
so $\sigma(\mu)$ and $\sigma_{\mathrm{ref}}$ (or $\sigma_{\mathrm{ref}}^{c}$)
pass through the same data-to-Monte-Carlo transition, and the ratio
$\varrho_\mu=\sigma(\mu)/\sigma_{\mathrm{ref}}$ is comparatively stable
in $\mathscr{L}$ (\cref{fig:lumi_scan}). Whether $\varrho_\mu$ drifts
towards or away from unity as $\mathscr{L}$ grows depends on whether the
signal-rich, high-score bins are well simulated (large $\tau_{ck}$) or
Monte-Carlo-starved (small $\tau_{ck}$); the \TREx\ \texttt{TransfoD}
binning rule ($z_b=z_s=10$, \cref{sec:statistics}) is designed to keep
$\tau_{ck}$ from becoming too small.
\end{appendices}
\end{document}